\documentclass[usenatbib]{mn2e}
\usepackage{natbibmnfix, graphicx, times, amsmath, epsfig, amssymb}

\newcommand{\cmfastSM}{\textsc{\small 21CMFASTv2}}
\newcommand{\cmfast}{\textsc{\small 21CMFAST}}

\newcommand{\avenf}{\bar{x}_{\rm HI}}

\newcommand{\lya}{Ly$\alpha$}

\newcommand{\Msun}{M_\odot}

\newcommand\lsim{\mathrel{\rlap{\lower4pt\hbox{\hskip1pt$\sim$}}
        \raise1pt\hbox{$<$}}}
\newcommand\gsim{\mathrel{\rlap{\lower4pt\hbox{\hskip1pt$\sim$}}
        \raise1pt\hbox{$>$}}}
\def\myputfigure#1#2#3#4#5%
{\vskip#5pt\makebox[0pt]{\hskip#2in
\includegraphics[width=#3\textwidth]{#1}}\vskip#4pt\hfill}

\pdfoutput=1

\begin{document}

\title[21cm -- LAE correlation]{Cross-correlation of the cosmic 21-cm signal and Lyman alpha emitters during reionization}
\author[Sobacchi, Mesinger, \& Greig]{Emanuele Sobacchi\thanks{email: emanuele.sobacchi@sns.it}, Andrei Mesinger, \& Bradley Greig  \\
Scuola Normale Superiore, Piazza dei Cavalieri 7, 56126 Pisa, Italy\\
}
\voffset-.6in

\maketitle

\begin{abstract}
  Interferometry of the cosmic 21-cm signal is set to revolutionize our understanding of the Epoch of Reionization (EoR), eventually providing 3D maps of the early Universe.  Initial detections however will be low signal-to-noise, limited by systematics.  To confirm a putative 21-cm detection, and check the accuracy of 21-cm data analysis pipelines, it would be very useful to cross-correlate against a genuine cosmological signal.  The most promising cosmological signals are wide-field maps of Lyman alpha emitting galaxies (LAEs), expected from the Subaru Hyper-Suprime Cam (HSC) Ultra-Deep field.
  Here we present estimates of the correlation between LAE maps at $z\sim7$ and the 21-cm signal observed by both the Low Frequency Array (LOFAR) and the planned Square Kilometer Array Phase 1 (SKA1).
  We adopt a systematic approach, varying both: (i) the prescription of assigning LAEs to host halos; and (ii) the large-scale structure of neutral and ionized regions (i.e. EoR morphology).
  We find that the LAE-21cm cross-correlation is insensitive to (i), thus making it a robust probe of the EoR.  A $1000$ h observation with LOFAR would be sufficient to discriminate at $\gsim 1 \sigma$ a fully ionized Universe from one with a mean neutral fraction of  $\bar{x}_{\rm HI}\approx0.50$, using the LAE-21cm cross-correlation function on scales of $R\approx$3--10 Mpc.
  Unlike LOFAR, whose detection of the LAE-21cm cross-correlation is limited by noise, SKA1 is mostly limited by ignorance of the EoR morphology.  However, the planned 100 h wide-field SKA1-Low survey will be sufficient to discriminate an ionized Universe from one with $\bar{x}_{\rm HI}=0.25$, even with maximally pessimistic assumptions.  
\end{abstract}

\begin{keywords}
cosmology: theory -- early Universe  -- dark ages, reionization, first stars -- galaxies: formation -- high-redshift -- evolution
\end{keywords}

\section{Introduction}

Although it is the last major phase change in the history of our Universe, the epoch of reionization (EoR) remains poorly explored.
The morphology and evolution of the EoR is driven by the birth, growth, and death of the first galaxies.  Observing this complex, physics-rich epoch requires significant observational efforts.
Arguably the most promising among upcoming EoR probes are: (i) wide-field Lyman alpha emitter (LAE) surveys, and in the long term, (ii) the cosmic 21-cm signal from neutral hydrogen.

Due to the high optical depth of the neutral intergalactic medium (IGM) to \lya\ radiation, the observed number of galaxies strongly emitting \lya\ is expected to drop rapidly during the EoR.
This effect has recently been tentatively confirmed by $z\gsim6$ observations of color-selected \lya\ emitting galaxies
(e.g. \citealt{Ouchi10, Fontana10, Stark10, Kashikawa11, Caruana12, Treu13, Konno14, Schenker14, Caruana14, Pentericci14, Cassata15}). However, it is difficult to disentangle a genuine EoR signature from an intrinsic evolution of galaxy properties (e.g. \citealt{DW10, DMF11, BH13, Jensen13, Dijkstra14, TL14, Mesinger15, Choudhury15}).  Moreover, current sample sizes are small enough to support a redshift evolution only at $\lsim 2\sigma$.  An alternative diagnostic is the clustering of observed LAEs, which is expected to increase with the cosmic neutral fraction $\bar{x}_{\rm HI}$ (e.g. \citealt{FZH06, McQuinn07LAE, MF08LAE, Jensen14}). However, since the dependence of the clustering on $\bar{x}_{\rm HI}$ is degenerate with the (unknown) typical mass of the host halos, current clustering measurements at $z\sim7$ from Subaru Suprime Cam can provide only an upper limit, $\bar{x}_{\rm HI}\lesssim$ 0.5 (0.65) at 1 (2) $\sigma$ (\citealt{SM15}; see also \citealt{McQuinn07LAE}).  Limits with the Hyper-Suprime Cam (HSC) upgrade are set to improve by a factor of few, though they are still strongly limited by the degeneracy between the intrinsic clustering of LAE host halos, and reionization induced clustering.
%$z\approx 6.6$even with the large volumes ($\sim 30\text{ deg$^2$}$ at $z=6.6$) probed by the upcoming surveys \citep{SM15}.

On the other hand, current $21\text{ cm}$ interferometers, such as the Low Frequency Array (LOFAR; \citealt{vanHaarlem13})\footnote{http://www.lofar.org/},  the Murchison Wide Field Array (MWA; \citealt{Tingay12})\footnote{http://www.mwatelescope.org/} and the Precision Array for Probing the Epoch of Reionization (PAPER; \citealt{Parsons10})\footnote{http://eor.berkeley.edu}, are aiming for a statistical detection of the EoR from the redshift evolution of large-scale 21-cm power.
Planned second-generation instruments, like the Square Kilometer Array (SKA)\footnote{https://www.skatelescope.org} and the Hydrogen Epoch of Reionization Array (HERA)\footnote{http://reionization.org} should even be able to provide the first tomographic maps of the 21-cm signal from the EoR, deepening our understanding of reionization-era physics.

With any 21-cm instrument, progress will be limited by systematics, such as improper calibration and modeling of the antennae response, the sky model, the ionosphere, foreground structure and radio frequency interference (e.g. \citealt{LPT14, Chapman14}).  Credible observations will rely on a constant reevaluation of our data analysis, adaptively improving our ability to deal with systematics in a slow march towards increasing signal to noise (S/N).  Thus initial claims of a detection will likely be met with skepticism and uncertainty as to whether the signal is genuinely of cosmic origins or is, for example, a foreground residual.

Therefore it would be very useful to cross-correlate such putative 21-cm detections with another cosmic signal.
An obvious choice for this are high-$z$ galaxy surveys (e.g. \citealt{FL07}).  Color-selected surveys are prone to large photometric redshift uncertainties.  In contrast, wide-field, narrow-band searches for LAEs are ideally suited for this cross-correlation in the short term (e.g. \citealt{PKWL14, Vrbanec15}).

Here we present estimates of the {\it cross-correlation of the upcoming Subaru HSC survey of $z=6.6$ LAEs and the cosmic 21-cm signal as observed by LOFAR and SKA}\footnote{Although set in the southern hemisphere, there is a reasonable overlap of the SKA and Subaru fields of view.  Unfortunately, drift-scan instruments such as PAPER and HERA would not be well suited for such a cross-correlation.}.  Our work is similar to the recent LOFAR-focused study of \citet{Vrbanec15}, which appeared as this work was nearing completion.  The notable differences are: (i) rather than using a single EoR and a single LAE model, we adopt a systematic approach by varying both the method to assign LAEs to host halos as well as the morphology and evolution of the EoR; and (ii) we also present forecasts for SKA.

This paper is organized as follows: in Section \ref{sec:methods} we describe the models for reionization and LAEs that we use to predict the cross-correlation statistics, as well as the assumed survey parameters.  In Section \ref{sec:results} we present our forecasts for both LOFAR and SKA.
In Section \ref{sec:concl} we present our conclusions. Throughout we assume a flat $\Lambda$CDM cosmology with parameters ($\Omega_{\rm m}$, $\Omega_{\rm \Lambda}$, $\Omega_{\rm b}$, $h$, $\sigma_{\rm 8}$, $n$) = ($0.28$, $0.72$, $0.046$, $0.70$, $0.82$, $0.96$),
consistent with recent results from the Planck satellite \citep{Planck15}. Unless stated otherwise, we quote all quantities in comoving units.

%%%%%%%%%%%%%%%%%%%%%%%%%%%%%%%%%%%%%%%%%%%%%%%%%%%%%%%%%%%%%%%%%%%%%%%%%%%%%%%%%%%%%%%%%%%%%%%%%%%%%%%%%%%%%%%%%%%%%%%%%%%%%%%%%%%%%%%%
%%%%%%%%%%%%%%%%%%%%%%%%%%%%%%%%%%%%%%%%%%%%%%%%%%%%%%%%%%%%%%%%%%%%%%%%%%%%%%%%%%%%%%%%%%%%%%%%%%%%%%%%%%%%%%%%%%%%%%%%%%%%%%%%%%%%%%%%
\section{Methods}
\label{sec:methods}

%%%%%%%%%%%%%%%%%%%%%%%%%%%%%%%%%%%%%%%%%%%%%%%%%%%%%%%%%%%%%%%%%%%%%%%%%%%%%%%%%%%%%%%%%%%%%%%%%%%%%%%%%%%%%%%%%%%%%%%%%%%%%%%%%%%%%%%%
\subsection{Theoretical Models}

Computing the cross-correlation of the 21-cm signal and the LAE field requires two components: (i) large-scale reionization simulations for determining the 21-cm brightness temperature, as well as the the \lya\ opacity of the IGM which attenuates the \lya\ line; and (ii) a scheme for assigning the intrinsic \lya\ luminosity (escaping the galaxy) to host DM halos.  As in \citet{SM15}, we adopt a systematic approach and explore extreme models for both (i) and (ii).  Below we describe each in turn, referring the reader to \citet{SM15} for more details.

\subsection{EoR simulations}

\begin{figure*}
\vspace{+0\baselineskip}
{
\includegraphics[width=0.45\textwidth]{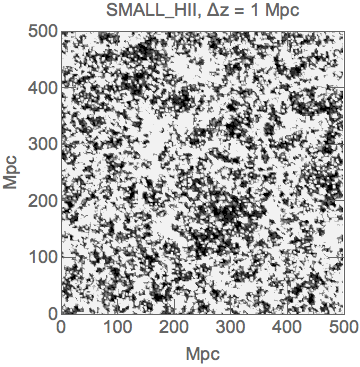}\qquad
\includegraphics[width=0.45\textwidth]{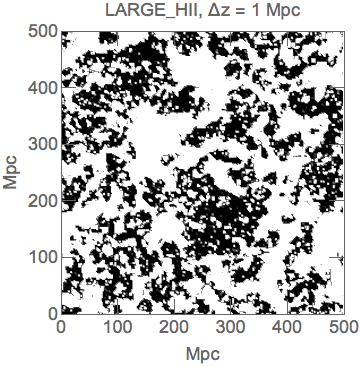}
\includegraphics[width=0.45\textwidth]{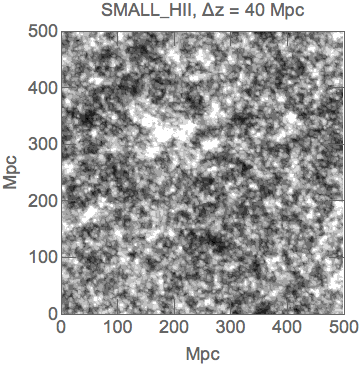}\qquad
\includegraphics[width=0.45\textwidth]{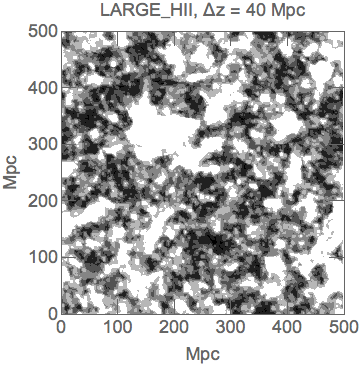}
\includegraphics[width=0.99\textwidth]{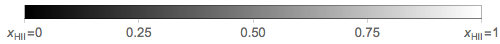}
}
\caption{Ionization fields (white/black pixels correspond to ionized/neutral regions) for the \textbf{SMALL\pmb{\_}HII} (left) and the \textbf{LARGE\pmb{\_}HII} (right) models. In the upper/lower panels we show a $1\text{ Mpc}$/$40\text{ Mpc}$ thick slice, corresponding to our grid resolution and to the HSC redshift uncertainty respectively. In the lower panels the superposition of different ionization structures along the line of sight causes some information loss on the ionization structure of the IGM.
\label{fig:reion}
}
\vspace{-1\baselineskip}
\end{figure*}

We model cosmological reionization using a modified version of the publically available code \cmfast\footnote{http://homepage.sns.it/mesinger/Sim.html} \citep{MF07, MFC11}.  Specifically, we use \cmfastSM\ \citep{SM14},  which incorporates calibrated, sub-grid prescription for inhomogeneous recombinations and photo-heating suppression of the gas fraction in small galaxies. Our boxes are $L=500\text{ Mpc}$ on a side with a final resolution of 500$^3$.  In its simplest variant, \cmfastSM\ has two free parameters: (i) the ionizing efficiency of star forming galaxies, $\zeta$; and (ii) the minimum mass scale above which SNe feedback is assumed to be inefficient at quenching star formation, $M^{\rm min}_{\rm SNe}$.\footnote{Note that the inhomogeneous photon mean free path and sub-grid recombinations are computed self consistently in \citet{SM14}, removing the need to impose a maximum photon horizon as is commonly done in \cmfast.  Additionally, we follow the inhomogeneous photo-heating suppression of the gas content of galaxies, approximating this as a sharp cut in the star formation threshold for halos with $M_{\rm h} > M_{\rm photo}$ calibrated to the simulation suites from \citet{SM13a}.}
(i)  impacts the timing of reionization, while (ii) additionally impacts the clustering of typical EoR sources and by extension the resulting EoR morphology.
%As in \citet{SM15}, we allow the timing of reionization to vary, keeping the value of $\avenf$ at $z=6.6$ a free parameter in our models.
As the cross-correlation of galaxies and 21-cm is sensitive to the EoR morphology, we explore two extreme models varying $M^{\rm min}_{\rm SNe}$:

\begin{itemize}
\item \textbf{SMALL\pmb{\_}HII}: here we assume that only the atomic cooling threshold, $M_{\rm h} > M^{\rm min}_{\rm cool}(T_{\rm vir}=10^4\text{ K})$, and photo-heating feedback, $M_{\rm h} > M^{\rm min}_{\rm photo}$ result in a strong suppression of star formation inside small mass halos.  The effects of SNe feedback are assumed to be halo mass independent above these thresholds, with no sharp suppression below a relevant mass scale (effectively, $M^{\rm min}_{\rm SNe} < \max[M_{\rm cool}, M_{\rm photo}]$).  Because the resulting ionizing sources are very weakly biased and susceptible to photo-heating feedback and its coupling to inhomogeneous recombinations (c.f. the ``FULL'' model in \citealt{SM14}), the resulting EoR morphology
is {\it characterized by small cosmic HII regions}, as seen in the top left panel of Fig. \ref{fig:reion}.\\

\item \textbf{LARGE\pmb{\_}HII}: here we assume that star-formation is efficient only in the rarest, most-massive, most-biased halos, taking the extreme value of $M^{\rm min}_{\rm SNe}(T_{\rm vir}=10^5\text{ K})$, approximately corresponding to the observed Lyman break galaxy candidates (LBGs; e.g. \citealt{KF-G12}).  The resulting EoR morphologies are
    {\it characterized by large cosmic HII patches}, as seen in the top right panel of Fig. \ref{fig:reion}.
\end{itemize}

These two models roughly bracket the uncertainty in the EoR morphology, at a fixed value of $\avenf$. As in \citet{SM15}, we allow the timing of reionization to vary, keeping $\avenf$ a free parameter in our models, and applying EoR maps corresponding to various values of $\avenf$ to our $z=6.6$ LAE fields (e.g. \citealt{McQuinn07, Jensen14}).

\subsection{Assigning LAEs to dark matter halos}

The cross-correlation will also depend on how the LAEs are clustered.  To assign \lya\ emission to host halos, we follow the same parametric approach as in \citet{SM15}:
The observed \lya\ luminosity is
\begin{equation}
L_{\alpha}=L_{\alpha}^{\rm intr}e^{-\tau_{\rm IGM}} ~,
\end{equation}
where $\tau_{\rm IGM}$ is the \lya\ optical depth along the line of sight. The observed LAEs in our mock surveys have $L_{\alpha}>2.5\times 10^{42}\text{ erg s$^{-1}$}$ corresponding to the Subaru UDFs with both the current SC and upcoming HSC surveys (M. Ouchi, private communication).  The intrinsic \lya\ luminosity is given by
\begin{equation}
L_{\alpha}^{\rm intr}=L_\alpha^{\rm min}\left(\frac{M_{\rm h}}{M_{\rm \alpha}^{\rm min}}\right)^\beta \chi ~ ,
\end{equation}
where $L_\alpha^{\rm min}$ is a normalization constant determined by the observed \lya\ luminosity function \citep{Ouchi10, Matthee15}, i.e. the luminosity corresponding to a galaxy residing in a halo of mass $M_{\rm \alpha}^{\rm min}$, and $\chi$ is a random variable ($\chi=1$ with probability $f_{\rm duty}$ and $\chi=0$ otherwise).  For a given choice of $M_{\rm \alpha}^{\rm min}$, we tune the \lya\ duty cycle, $f_{\rm duty}$, to match the {\it observed} (i.e. post IGM attenuation) number density of $z=6.6$ LAEs, $\bar{n}_{\rm LAE}=4.8\times 10^{-4}\text{ Mpc$^{-3}$}$ \citep{Ouchi10}.

Consistent with $z\sim4$ LAEs, we take $\beta=1$ above (e.g. \citealt{Gronke15}), and we also evaluate the IGM opacity at a typical velocity offset of $\Delta v_{\rm sys} =$200 km s$^{-1}$ from the systemic redshift (e.g. \citealt{Shibuya14, Stark15, Sobral15}).  We stress however that the clustering, normalized to a fixed number density, is extremely insensitive to these choices (see the appendix of \citealt{SM15}).

%Combining this abundance matching and current LAE clustering measurements at $z=6.6$ \citep{Ouchi10} constrains $f_{\rm duty}\lesssim\text{ few \%}$, which corresponds to an average mass $\bar{M}_{\rm h}\lesssim 2\times 10^{10}M_\odot$ for the host DM halos \citep{SM15}. Thus,
To bracket the allowed range for the intrinsic clustering of LAEs, we consider two models varying the minimum host halo mass, $M_{\rm \alpha}^{\rm min}$:
\begin{itemize}
\item {\bf Massive halos} --  this model results in an average halo mass hosting LAEs of {\bf $\bar{M}_{\rm h} \approx 2 \times 10^{10} \Msun$},
    chosen to maximize the intrinsic clustering of $z=6.6$ LAEs, within the current observational limits \citep{SM15}.
  The corresponding duty cycle is $f_{\rm duty}\approx 0.02$, assuming a mostly-ionized Universe.\\

\item {\bf Low mass halos} --  this model results in an average halo mass hosting LAEs of {\bf $\bar{M}_{\rm h} \approx 3 \times 10^{9} \Msun$},
    chosen to minimize the intrinsic clustering of $z=6.6$ LAEs. The corresponding duty cycle is $f_{\rm duty}\approx 0.001$, assuming a mostly-ionized Universe.
\end{itemize}

After assigning intrinsic \lya\ luminosities and computing the IGM optical depths, 
the mock LAE map is constructed by cutting the simulation box into slabs of width $\simeq 40\text{ Mpc}$, corresponding to the redshift uncertainty, $\Delta z\simeq 0.1$, for the narrow-band LAE surveys.  We also remove the external cells to match the survey area, $\sim 0.1\text{ Gpc$^2$}$.

\subsection{Cross-correlation statistics}

To quantify the cross-correlation of the 21$\text{ cm}$ signal and the galaxy maps, we use two different statistics: the cross-correlation coefficient (CCC) and the real space cross-correlation function (RSCF). Starting from our simulated galaxy and $21\text{ cm}$ maps, we smooth the galaxy distribution on our grid, calculating the galaxy overdensity field:
\begin{equation}
\delta_{\rm gal}\left(\textbf{x}\right)=\frac{N_{\rm gal}\left(\textbf{x}\right)}{\bar{N}_{\rm gal}}-1~,
\end{equation}
where $N_{\rm gal}\left(\textbf{x}\right)$ is the number of galaxies in the voxel and $\bar{N}_{\rm gal}$ is the average number. We use the non-dimensional $21\text{ cm}$ brightness temperature defined as
\begin{equation}
\delta_{\rm 21}\left(\textbf{x}\right)=\frac{T_{\rm 21}\left(\textbf{x}\right)-\bar{T}_{\rm 21}}{T_{\rm 0}} ~,
\end{equation}
where $\bar{T}_{\rm 21}$ is the average temperature and the normalization $T_{\rm 0}=23.5\text{ mK}$ is the expected brightness temperature at $z=6.6$ if the Universe is entirely neutral.

The cross correlation coefficient is defined as
\begin{equation}
CCC_{\rm 21, gal}\left(k\right)= \frac{P_{\rm 21, gal}\left(k\right)}{\sqrt{P_{\rm 21}\left(k\right)P_{\rm gal}\left(k\right)}}~,
\end{equation}
where $P_{\rm 21}\equiv k^3/(2\pi^2 V) ~ \langle|\delta_{\rm 21}|^2\rangle_k$ and $P_{\rm gal}\equiv k^3/(2\pi^2 V) ~ \langle|\delta_{\rm gal}|^2\rangle_k$ are the usual $21\text{ cm}$ and galaxies power spectra, while $P_{\rm 21, gal}\equiv k^3/(2\pi^2 V) ~ \Re\langle\delta_{\rm 21}\bar{\delta}_{\rm gal}\rangle_k$ is the cross-power spectrum. The CCC can then be understood as the expectation value of the cosine of the phase difference between $\delta_{\rm gal}\left(\textbf{k}\right)$ and $\delta_{\rm 21}\left(\textbf{k}\right)$:
%AM: maybe write this one step out further, expanding eq. 5 to a new line to show the expectation value of the phase diff
%ES: Done. Please check if the following equation (new eq. 6) is what you were thinking at.
\begin{eqnarray}
CCC_{\rm 21, gal}\left(k\right) & = & \frac{\Re\langle|\delta_{\rm 21}|\text{e}^{i\theta_{\rm 21}}|\delta_{\rm gal}|\text{e}^{-i\theta_{\rm gal}}\rangle_k}{\sqrt{\langle|\delta_{\rm 21}|^2\rangle_k\langle|\delta_{\rm gal}|^2\rangle_k}}= \nonumber\\
& = & \Re\langle\text{e}^{i\theta_{\rm 21}-i\theta_{\rm gal}}\rangle_k=\langle\cos\left(\theta_{\rm 21}-\theta_{\rm gal}\right)\rangle_k~.
\end{eqnarray}
If the modes are strongly correlated, $CCC\rightarrow1$; if the the modes are strongly anti-correlated, $CCC \rightarrow -1$; while if the phase shift is random with no correlation, the $CCC$ will average to zero for those modes.

The cross-correlation function $r_{\rm 21, gal}\left(r\right)\equiv \langle\delta_{\rm 21}\left(x\right)\delta_{\rm gal}\left(x+r\right)\rangle_x$ is a more commonly used statistic.  Since it is also a probability in excess of random, its amplitude is more physically intuitive than that of the CCC, as we shall see below.  For 3D fields, we compute the RSCF as the fourier transform of the cross-power spectrum $P_{\rm 21, gal}\left(k\right)$. However, for 2D maps we use the  real-space metric from \citet{Croft15}, which results in smaller noise. Namely, we
sum over all visible galaxy-21$\text{ cm}$ pixel pairs separated by a distance $r$:
(e.g. \citealt{Croft15}): 
\begin{equation}
r_{\rm 21, gal}\left(r\right)=\frac{1}{N_{\rm gal}N\left(r\right)}\sum_{i=1}^{N_{\rm gal}}\sum_{j=1}^{N\left(r\right)}\delta_{\rm 21}\left(r\right)~,
\label{eq:cross}
\end{equation}
where $N_{\rm gal}$ is the number of galaxies in the survey and $N\left(r\right)$ is the number of pixels at distance $r$ from the i-th galaxy.

%%%%%%%%%%%%%%%%%%%%%%%%%%%%%%%%%%%%%%%%%%%%%%%%%%%%%%%%%%%%%%%%%%%%%%%%%%%%%%%%%%%%%%%%%%%%%%%%%%%%%%%%%%%%%%%%%%%%%%%%%%%%%%%%%%%%%%%%
\subsection{Observational programs}
\label{sec:obs}

\subsubsection{HSC Ultra-Deep Field}
\label{sec:obs_LAE}

We model our mock LAE surveys on the basis of the planned UD campaign with the Subaru Hyper-Suprime Cam.  The UD field is probing $\sim4\text{ deg$^2$}$ (corresponding to $\sim 0.1\text{ Gpc$^2$}$) at $z=6.6$, likely large enough to allow us to statistically sample the EoR morphology (e.g. \citealt{Iliev14}; Mesinger et al., in prep). With a luminosity threshold $L_\alpha^{\rm min}=2.5\times 10^{42}\text{ erg s$^{-1}$}$, the expected number density of the observed LAEs is $\bar{n}_{\rm LAE}=4.8\times 10^{-4}\text{ Mpc$^{-3}$}$; the survey is going to have a systemic redshift uncertainty of $\Delta z=0.1$, corresponding to $\sim 40\text{ Mpc}$ at the redshift of the survey (M. Ouchi, private communication).

\subsubsection{LOFAR and SKA1-Low specifications}
\label{sec:obs_21cm}

\begin{table}
\centering
\caption{Summary of telescope parameters.  Parameters inside brackets correspond to the core stations for SKA1-Low, which will dominate the sensitivity for EoR measurements.}
\label{tab:21cm}
\begin{tabular}{l|c||r}
\hline
Parameter & LOFAR & SKA1-Low\\
\hline
Telescope antennae & 48 & 564 (240)\\
Diameter ($\text{m}$) & 30.8 & 30\\
Collecting area ($\text{m$^2$}$) & 35762 & 398668 (169646)\\
$T_{\rm rec}$ ($\text{K}$) & 140 & $0.1T_{\rm sky}+40$\\
Bandwidth ($\text{MHz}$) & 8 & 8\\
Integration time ($\text{h}$) & 1000 & 100-1000\\
\hline
\end{tabular}
\end{table}

Within this work, we restrict our analysis to two telescopes (LOFAR and SKA1-Low). In this section, we outline the specifics and assumptions we make in producing our telescope noise profiles, summarizing the key telescope parameters in Table \ref{tab:21cm}, and defer the reader to the more detailed discussions within \citet{Parsons12, Pober13, Pober14}. The thermal noise power spectrum is computed at each cell according to the following (e.g. \citealt{Morales05, McQuinn06, Pober14}):
\begin{equation}
\Delta_{\rm N}^2\left(k\right)=X^2Y\frac{k^3}{2\pi^2}\frac{\Omega'}{2t}T_{\rm sys}^2
\label{eq:21cm_noise}
\end{equation}
where $X^2Y$ is a cosmological conversion factor between observing bandwidth, frequency and comoving distance units, $\Omega'$ is a beam-dependent factor derived in \citet{Parsons14}, $t$ is the total time spent by all baselines within a particular $k$ mode and $T_{\rm sys}$ is the system temperature, the sum of the receiver temperature, $T_{\rm rec}$, and the sky temperature, $T_{\rm sky}$. For all telescope configurations considered in this work we assume tracked scanning with a total synthesis time of $6\text{ h}$ per night. We model $T_{\rm sky}$ using the frequency dependent scaling $T_{\rm sky}=60\left(\frac{\nu}{300\text{ MHz}}\right)^{-2.55}\text{ K}$ \citep{TMS07}.

We model LOFAR using the antennae positions listed in \citet{vanHaarlem13} and we assume $T_{\rm rec}=140\text{ K}$, consistent with \citet{Jensen13}. For SKA, we mimic the latest design configuration (V4A) for the SKA-low Phase 1 instrument outlined in the SKA1-Low configuration document;\footnote{http://astronomers.skatelescope.org/wp-content/uploads/2015/11/SKA1-Low-Configuration\_V4a.pdf} the total SKA system temperature is modelled as outlined in the SKA System Baseline Design, $T_{\rm sys} = 1.1T_{\rm sky} + 40\text{ K}$.
We compare two different methods to treat the foregrounds: in the fiducial one (foreground avoidance) we assume no foreground subtraction; this results in an extended $k$-space region (the ``wedge'') where the foregrounds completely dominate the $21\text{ cm}$ signal. In the optimistic one (foreground removal), the foregrounds are subtracted, including into the noise dominated "wedge", assuming an efficient cleaning algorithm; for more details on these assumptions, see \citet{Pober14}.

%%%%%%%%%%%%%%%%%%%%%%%%%%%%%%%%%%%%%%%%%%%%%%%%%%%%%%%%%%%%%%%%%%%%%%%%%%%%%%%%%%%%%%%%%%%%%%%%%%%%%%%%%%%%%%%%%%%%%%%%%%%%%%%%%%%%%%%%
%%%%%%%%%%%%%%%%%%%%%%%%%%%%%%%%%%%%%%%%%%%%%%%%%%%%%%%%%%%%%%%%%%%%%%%%%%%%%%%%%%%%%%%%%%%%%%%%%%%%%%%%%%%%%%%%%%%%%%%%%%%%%%%%%%%%%%%%
\section{Results}
\label{sec:results}

\subsection{Building physical intuition: halo--21 cm cross correlation}

\begin{figure*}
\vspace{+0\baselineskip}
{
\includegraphics[width=0.3\textwidth,height=3.8cm]{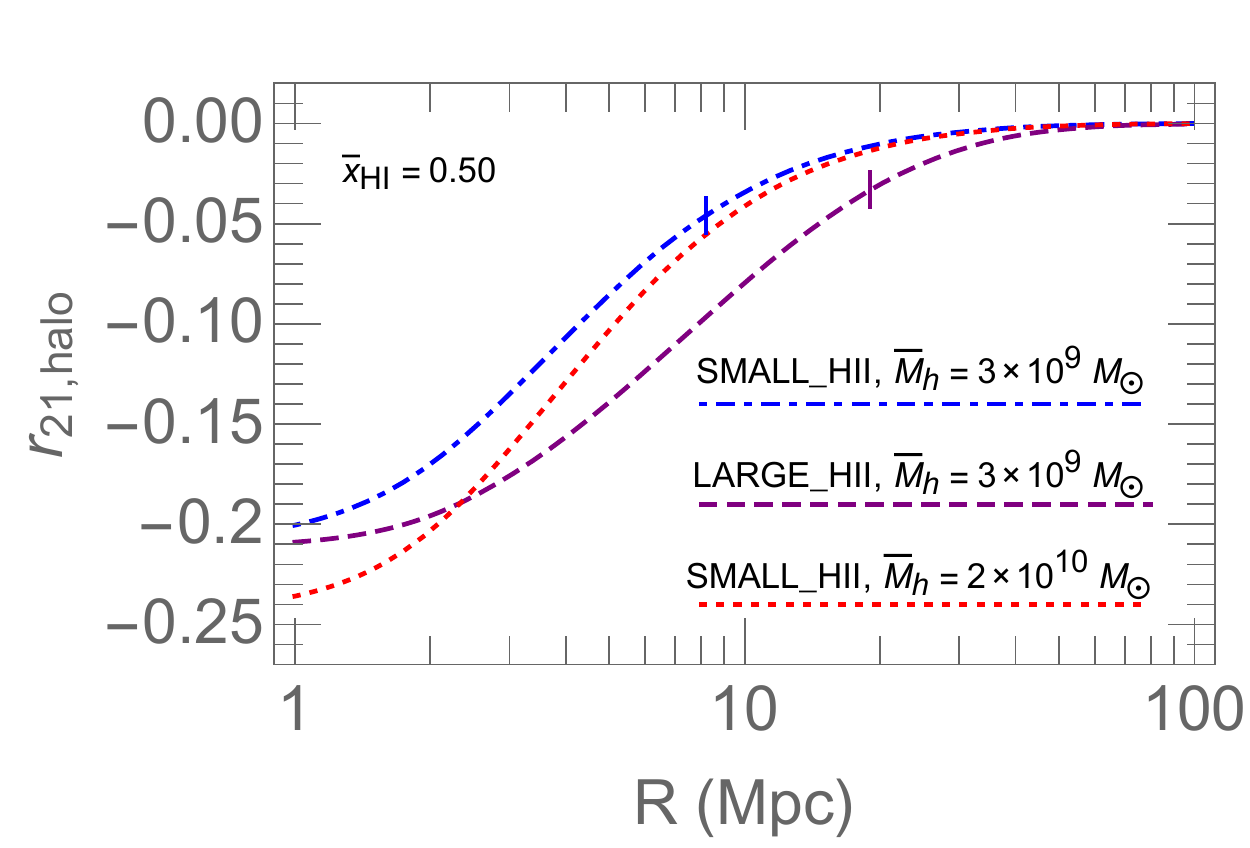}\qquad
\includegraphics[width=0.3\textwidth]{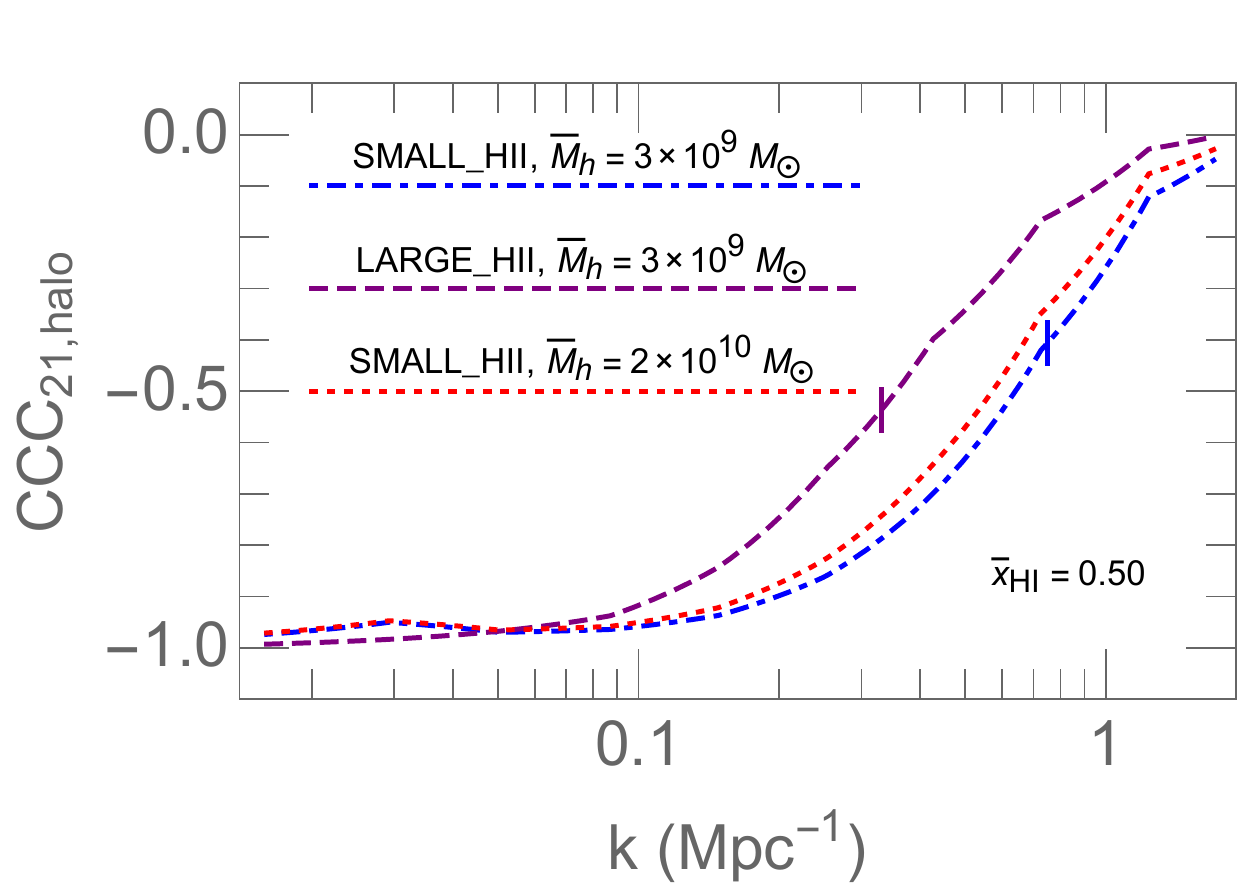}\qquad
\includegraphics[width=0.3\textwidth]{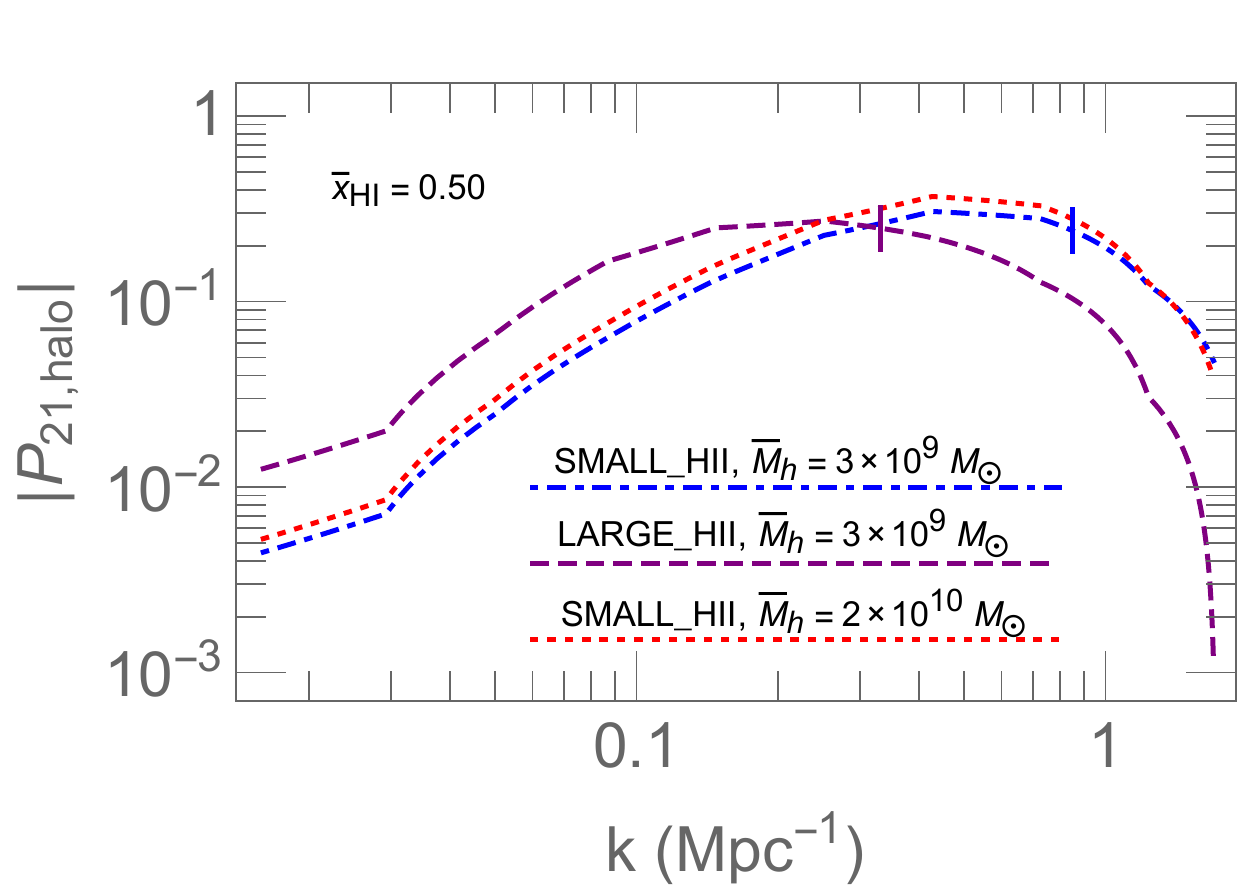}
}
\caption{Halo-$21\text{ cm}$ cross correlation function (left), cross correlation coefficient (middle) and cross-power spectrum (right), in 3D, for different reionization and halo models: \textbf{SMALL\pmb{\_}HII}, $\bar{M}_{\rm h}=3\times 10^{9}M_\odot$ (dot-dashed),  \textbf{LARGE\pmb{\_}HII}, $\bar{M}_{\rm h}=3\times 10^{9}M_\odot$ (dashed),  \textbf{SMALL\pmb{\_}HII}, $\bar{M}_{\rm h}=2\times 10^{10}M_\odot$ (dotted). The average HII region scale for the two EoR morphologies are denoted with vertical ticks on the panels.
\label{fig:3D}
}
\vspace{-1\baselineskip}
\end{figure*}
%AM: please cut the k plots at k~1--2 to skip the noise dominates small scales
%ES: Done

\begin{figure*}
\vspace{+0\baselineskip}
{
\includegraphics[width=0.3\textwidth,height=3.8cm]{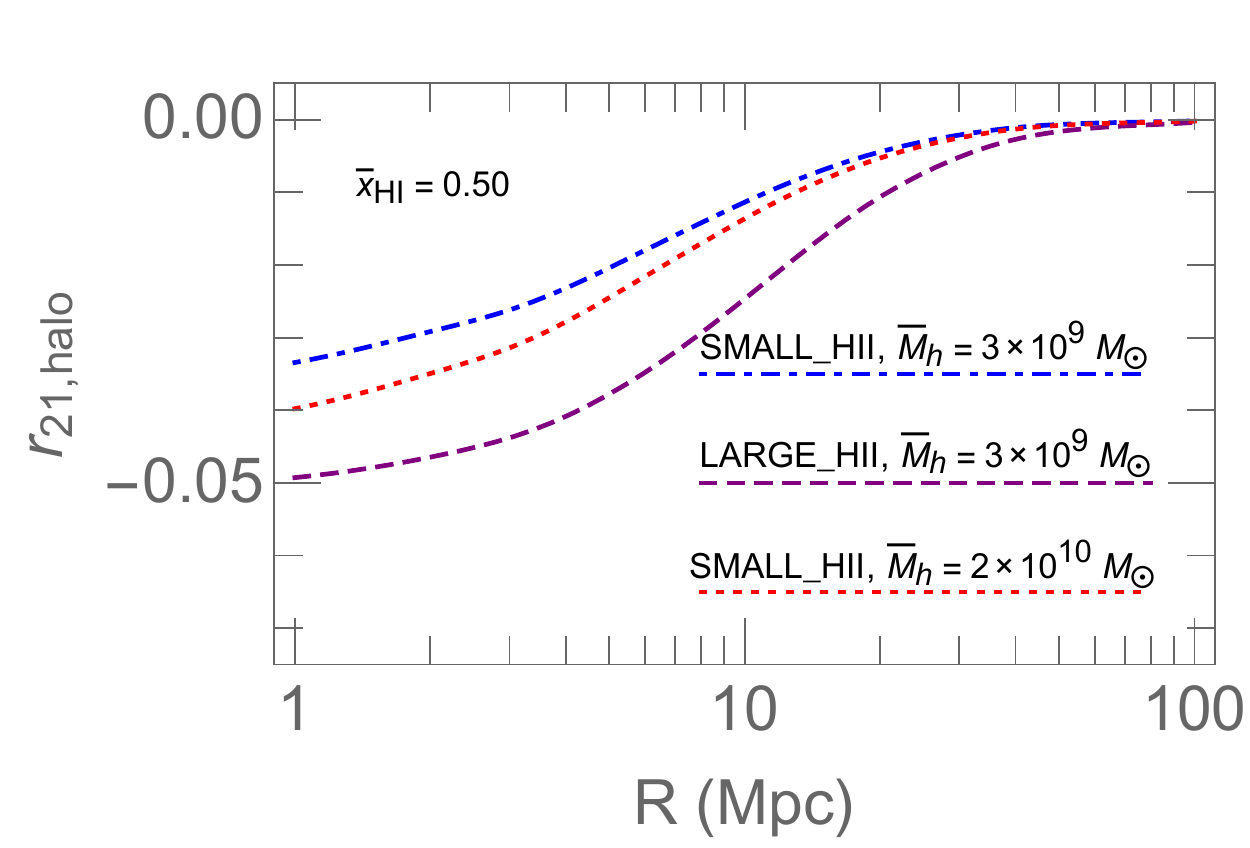}\qquad
\includegraphics[width=0.3\textwidth]{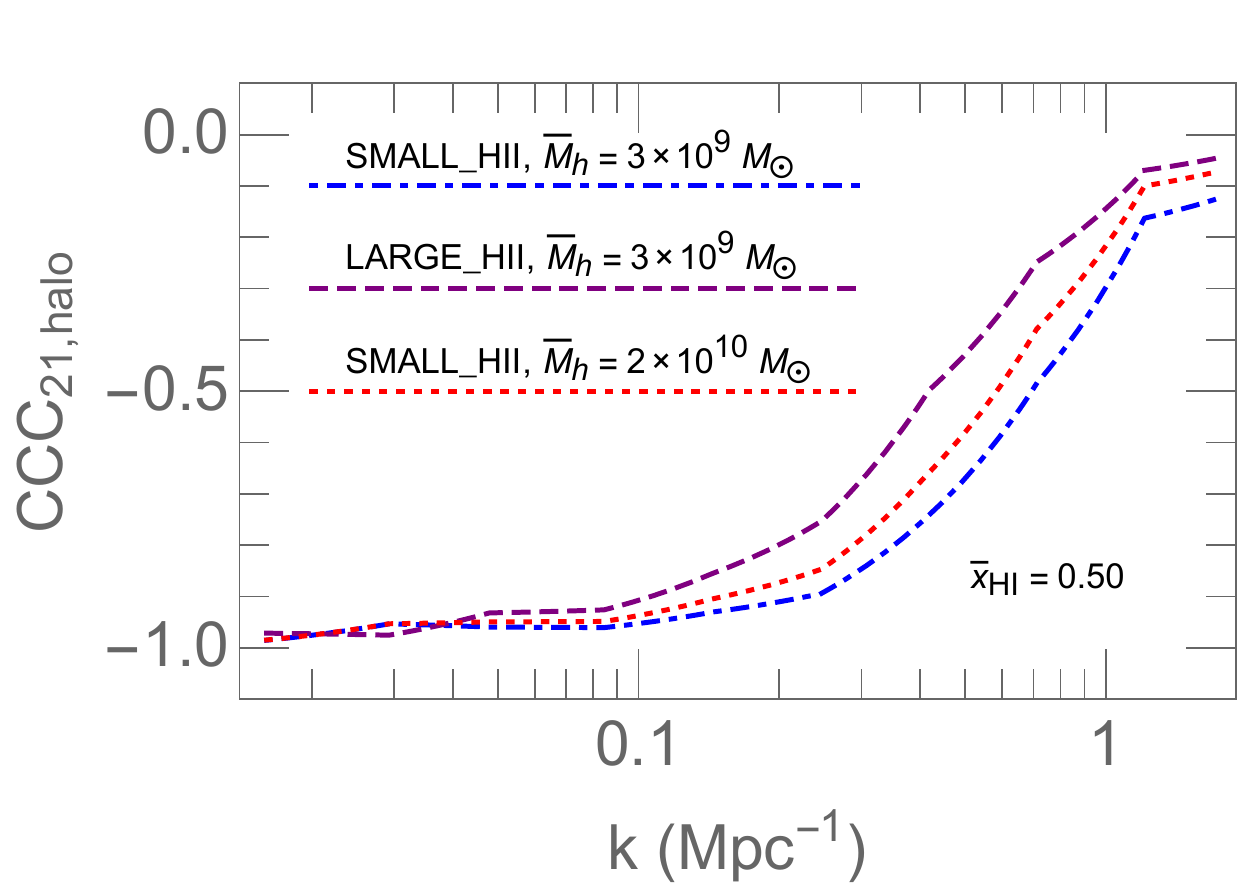}\qquad
\includegraphics[width=0.3\textwidth]{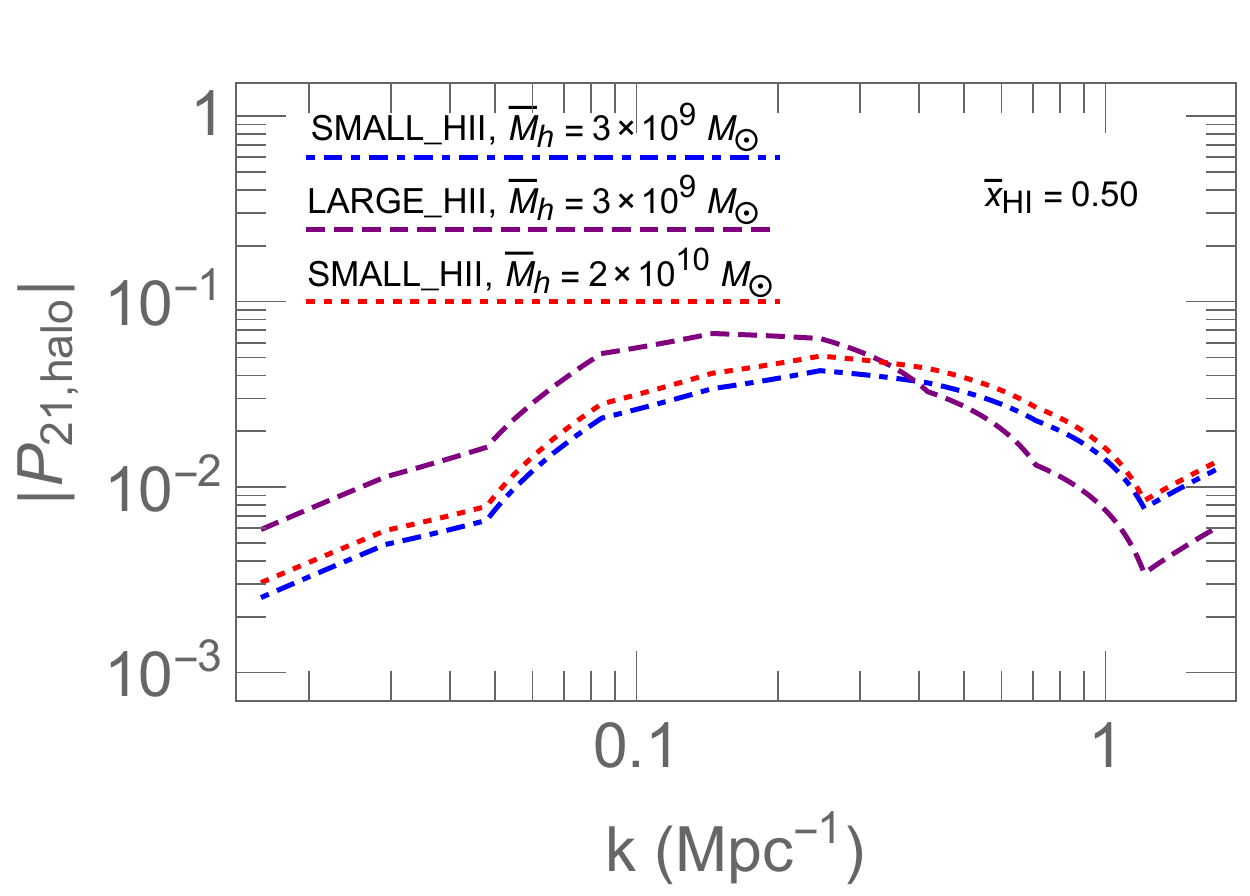}
}
\caption{Halo-$21\text{ cm}$ cross correlation function (left), cross correlation coefficient (middle) and cross-power spectrum (right), in 2D (projected over $\Delta z=0.1$), for different reionization and halo models: \textbf{SMALL\pmb{\_}HII}, $\bar{M}_{\rm h}=3\times 10^{9}M_\odot$ (dot-dashed),  \textbf{LARGE\pmb{\_}HII}, $\bar{M}_{\rm h}=3\times 10^{9}M_\odot$ (dashed),  \textbf{SMALL\pmb{\_}HII}, $\bar{M}_{\rm h}=2\times 10^{10}M_\odot$ (dotted).
\label{fig:2D}
}
\vspace{-1\baselineskip}
\end{figure*}
%AM: please remove the ticks in the 2D plots
%ES: Done

We begin by building physical intuition about the behavior of the CCCs and the RSCFs, studying their dependence on the underlying halo population and on the EoR morphology in the absence of noise.
In Fig. \ref{fig:3D} we show the 3D\footnote{Here we effectively assume that both the 21-cm signal and the LAEs can be localized on our native simulation grid, with 1 Mpc cells, without considering any further smoothing (c.f. the top panels in Fig. \ref{fig:reion}). We begin with the 3D fields as they are more intuitive, but note that measurements close to this level of resolution might be achievable in the future with: (i) spectroscopic confirmations of narrow-band selected LAEs (which dramatically reduce their systemic redshift uncertainties), combined with (ii) high S/N imaging with the SKA.}
halo-$21\text{ cm}$ RSCF (left), CCC (middle) and cross-power spectrum (right) at $\bar{x}_{\rm HI}=0.50$.
We vary both the EoR morphology and the average halo masses, as shown in the figure labels.  The average scale of HII regions, computed with the Monte-Carlo mean free path approach of \citet{MF07}, are denoted with vertical ticks on the panels.

The 3D RSCFs (left panel) show a clear negative correlation at small scales, as expected since inside HII regions we have $\delta_{\rm 21}<0$ and $\delta_{\rm halo}>0$, and a weak positive correlation at scales larger than the bubble size (e.g. \citealt{Lidz09,PKWL14,Vrbanec15}).  The turnover in the RSCFs shifts to larger scales if the EoR morphology is characterized by larger HII structures (c.f. blue and purple curves). The EoR morphology dominates the RSCFs on moderate to large scales ($R \sim$ 10 Mpc), while the intrinsic clustering of the halos dominates on scales smaller that typical HII sizes ($R  \sim 1$ Mpc).  

The CCCs (middle panel) show the same trends as the RSCF, with the turnover shifting to larger scales (smaller $k$) when the halo mass or the bubble size increases. However, there is a qualitative difference between the CCC and the RSCF: the RSCF is negative on small spatial scales, while the CCC is negative at small $k$. These seemingly opposite trends can be understood within the mindset of the peak-background split description of halo clustering, with the halo field being comprised of long and short wavelength modes which are only weakly correlated  (e.g. \citealt{CK89}).  Modes of the halo field with wavelengths much smaller than the typical HII bubble size can have their phases randomized, without impacting the cross-correlation (i.e. the halos will still sit in regions of zero 21-cm signal; \citealt{Lidz09}).  Thus, the cross-power and the CCCs drop to $\sim 0$ at large $k$.  However, both the halos (peaks of the $\delta_{\rm halo}$ field) and the HII regions (minima of the $\delta_{21}$ field) are sourced by density peaks on large-scales: 
thus they have a constant phase difference of $\sim\pi$ and the CCC is $\sim -1$ at small $k$.  Note that for very small $k$ ($\lesssim 0.1\text{ Mpc$^{-1}$}$), the amplitude of the cross-power drops rapidly (right panel), but the CCCs still have a value of $\sim -1$ as they {\it only measure the phase difference, not the amplitude of the cross power}.
On the other hand, the RSCFs, being the fourier transform of the cross power, involve a {\it weighted integral of the cross-power spectra over all $k$}.  Thus the cross-correlation seen on small scales in the RSCFs, is actually sourced by a wide range of moderate-$k$, where the cross-power amplitude is high.\footnote{For example, $\sim 90$\% of the value of the RSCF at $R=10\text{ Mpc}$ for our \textbf{SMALL\pmb{\_}HII}, $\bar{M}_{\rm h}=3\times 10^9M_\odot$ model is sourced by modes within $0.05\text{ Mpc$^{-1}$}<k<0.2\text{ Mpc$^{-1}$}$.}
%AM: please quantify this above when you compute the cross power.  you can even look at which k modes are dominant in setting the RSCFs at R=10, and plot this as a CDF...  i don't know if you have done this already, but i think it would be enough to say something like ``For example, blah\% of the value of the RSCF at R=10 for blah model is sourced by modes within $blah<k<blah''.  if this would take too much time, then never mind
%ES: I did this for R=10Mpc.
Going to larger spatial scales, the RSCFs drops, since the amplitude of the cross-power also drops at small $k$.

%Since real observations involve 2D fields, we consider the effect of projecting the ionization field along the line of sight. In Figure \ref{fig:reion} we show the $x_{\rm HI}$ fields for the \textbf{SMALL\pmb{\_}HII} (left) and the \textbf{LARGE\pmb{\_}HII} (right) models. In the upper panels we show a $1\text{ Mpc}$ thick slice (corresponding to our grid resolution); in the lower panels we show the neutral fraction smoothed on a $40\text{ Mpc}$ thick slice (corresponding to the HSC redshift uncertainty). By comparing the differences between the upper and lower panels, we see that the redshift uncertainties of $\Delta z \sim 0.1$ characteristic of narrow-band surveys result in a smoothing of EoR morphology and a subsequent loss of information.  As quantified in \citet{SM15}, spectroscopic confirmations would substantially decrease the uncertainty in the systemic galaxy redshifts, allowing us to more easily distinguish different EoR morphologies from the clustering of LAEs.

%The loss of information from systemic redshift uncertainties can be quantified by comparing the 3D and 2D correlation functions.

We now project the halo and 21-cm fields to 2D, within a thickness of $\Delta z=0.1$ characteristic of systemic redshift uncertainties of narrow-band surveys.  By comparing the differences between the upper and lower panels of Figure \ref{fig:reion}, we see that such systemic redshift uncertainties smear-out the ionization structure and the associated LAE clustering signature (e.g. \citealt{SM15}).  This loss of information is quantified in Fig. \ref{fig:2D}, showing the analogous quantities from the previous figure, but instead computed from the 2D fields.
The trends discussed above are preserved in 2D.  However the small scale behavior of the RSCF is washed out, since we are losing information on $k$-parallel modes up to scales of $\sim 40\text{ Mpc}$.  As discussed above, modes of roughly these scales dominate the anti-correlation signal on small scales in real space.  The CCCs are less affected in 2D as the mode mixing is less of an issue in $k$-space and the CCCs do not depend on the actual {\it amplitude} of the cross-power, which does decrease significantly going from 3D to 2D (right panels).

%%%%%%%%%%%%%%%%%%%%%%%%%%%%%%%%%%%%%%%%%%%%%%%%%%%%%%%%%%%%%%%%%%%%%%%%%%%%%%%%%%%%%%%%%%%%%%%%%%%%%%%%%%%%%%%%%%%%%%%%%%%%%%%%%%%%%%%%
\subsection{Realistic forecasts: LAE--21 cm cross correlation}

Now that we have explored the trends of the cross-correlation, we include the uncertainties expected from the upcoming surveys, as discussed in \S \ref{sec:obs}.  From now on, we focus on the RSCF statistic, for several reasons: (i)  as seen in the previous section, the CCCs contain analogous information on the cross-correlation, and so including both statistics would complicate the presentation without adding much additional information; (ii) as they correspond to an excess probability, we find the RSCFs to be more physically intuitive; and (iii) the RSCF is more robust to cosmic variance uncertainties, since it involves an integral over a broad range of moderate $k$-modes in the cross-power (which dominate the anti-correlation), rather than being computed in discrete, poorly-sampled $k$-bins like the CCCs.  The latter is especially important given the small duty cycles of LAEs, which result in a sizable sample variance at small $k$.

Below we present forecast for both LOFAR and SKA.  To bracket the likely signal, we calculate the RSCFs for all four combinations of our EoR morphologies and LAE models.  For each model, the uncertainty in the cross-correlation is computed from $10$ mock observations, which include both interferometer noise and sample variance.

\subsubsection{LOFAR}

\begin{figure*}
\vspace{+0\baselineskip}
{
\includegraphics[width=0.45\textwidth]{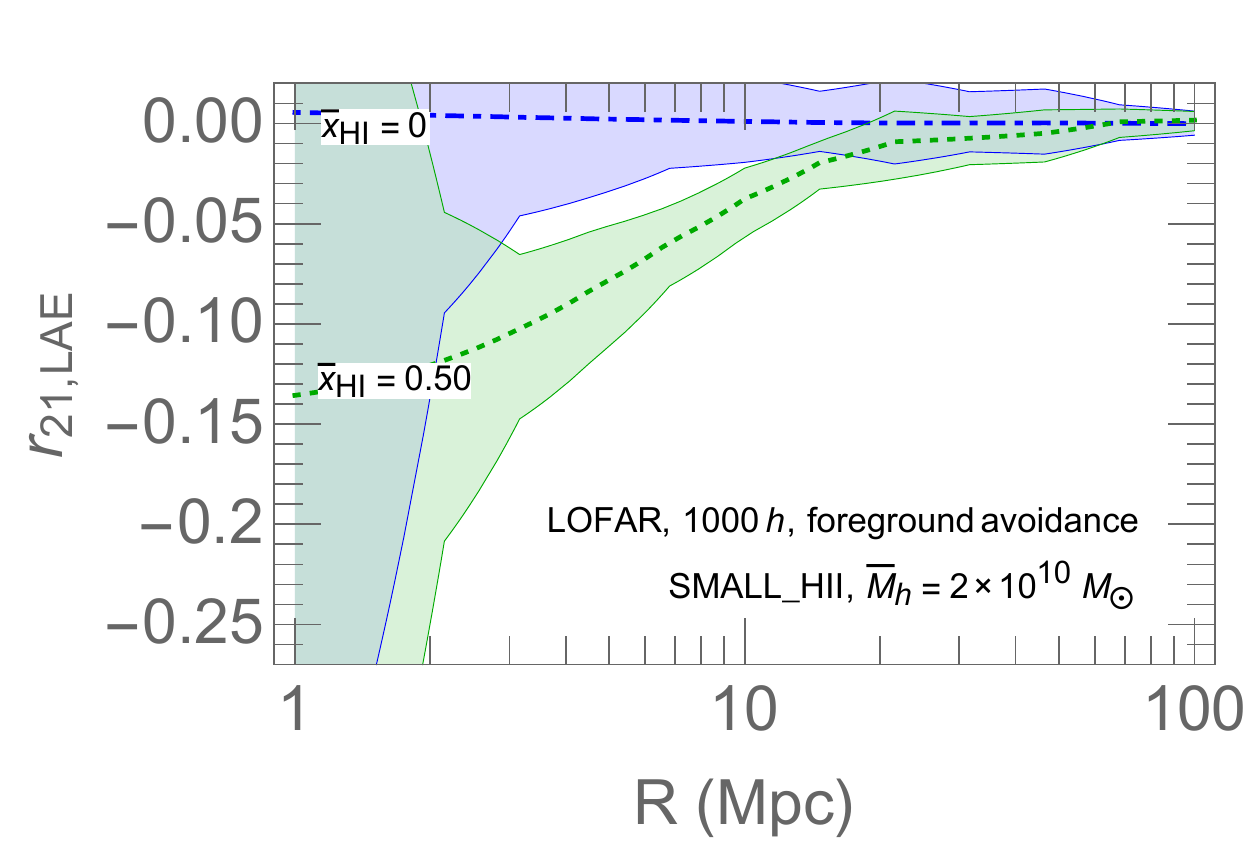}\qquad
\includegraphics[width=0.45\textwidth]{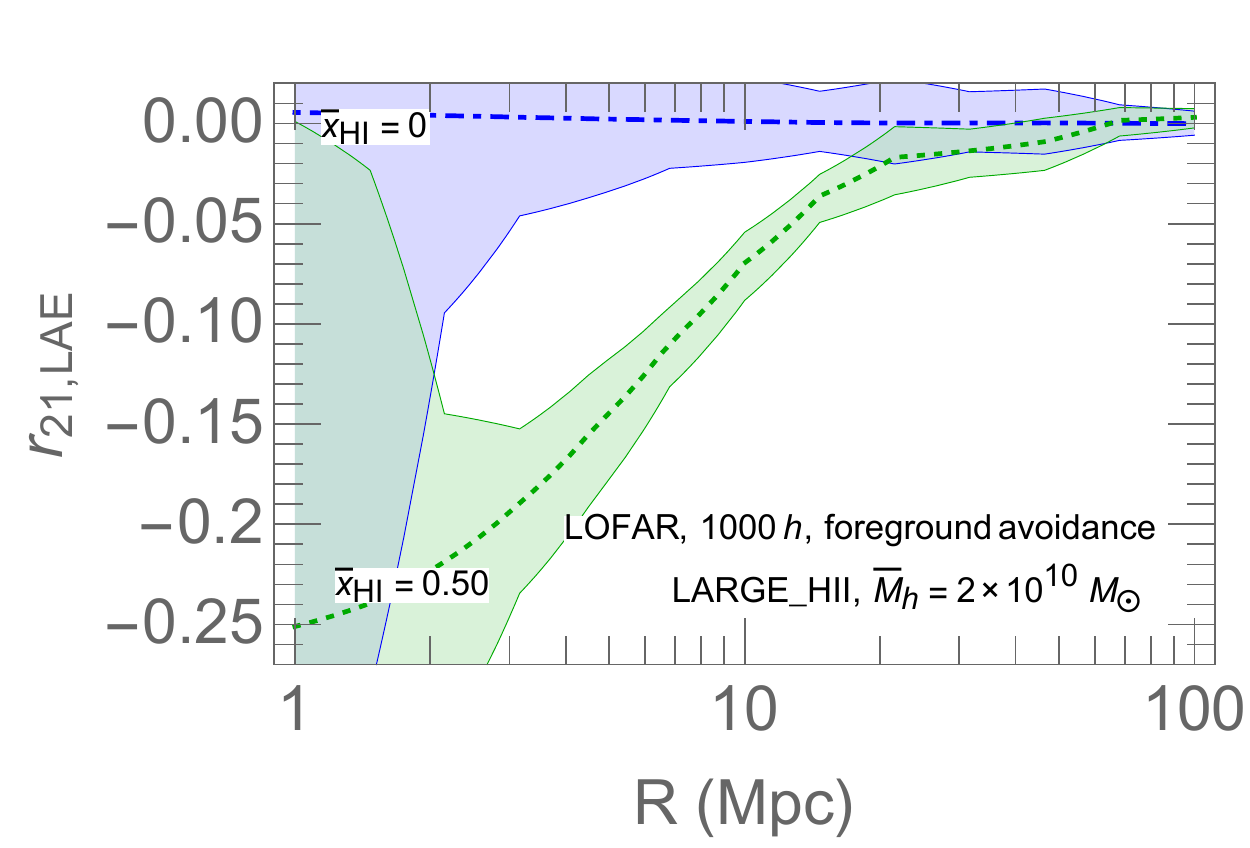}
\includegraphics[width=0.45\textwidth]{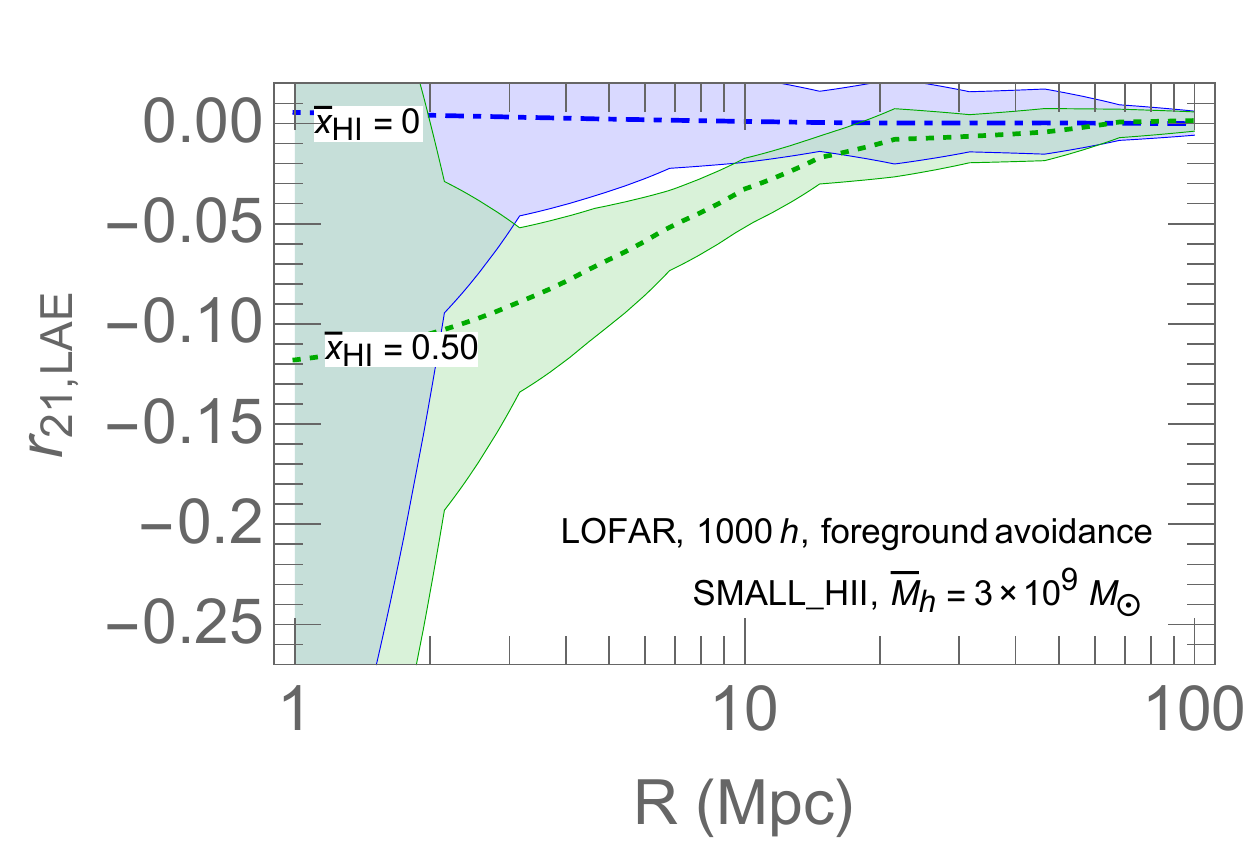}\qquad
\includegraphics[width=0.45\textwidth]{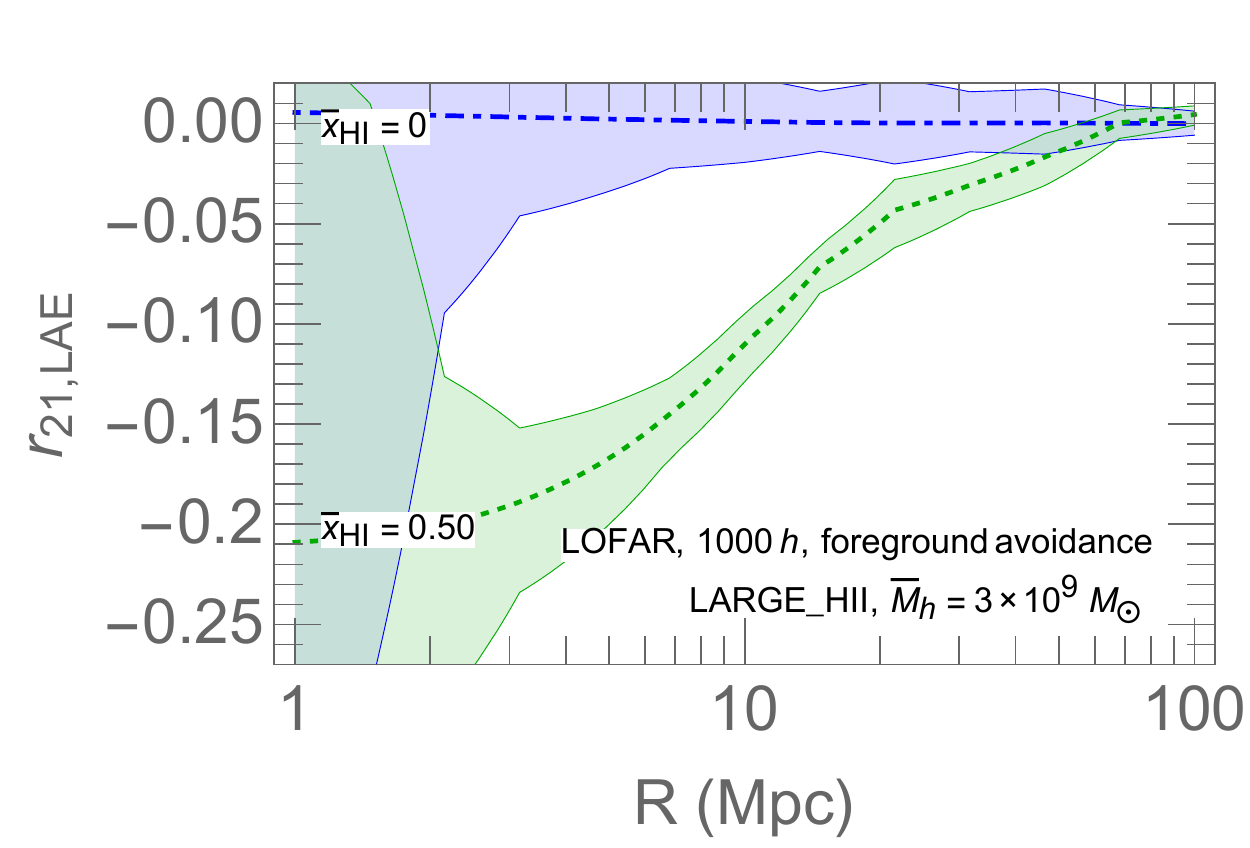}
}
\caption{LAE-$21\text{ cm}$ cross-correlation function at different average neutral fractions $\bar{x}_{\rm HI}=0$ (dot-dashed) and $\bar{x}_{\rm HI}=0.50$ (dotted). Shaded regions correspond to LOFAR noise plus sample variance for a $1000\text{ h}$ observation, with our conservative model for the foregrounds. We show different models for reionization (\textbf{SMALL\pmb{\_}HII}/\textbf{LARGE\pmb{\_}HII} in the left/right panels) and for the host halo masses ($\bar{M}_{\rm h}=2\times 10^{10}$/$3\times 10^9M_\odot$ in the upper/lower panels).
\label{fig:r_lofar_avo}
}
\vspace{-1\baselineskip}
\end{figure*}

\begin{figure*}
\vspace{+0\baselineskip}
{
\includegraphics[width=0.45\textwidth]{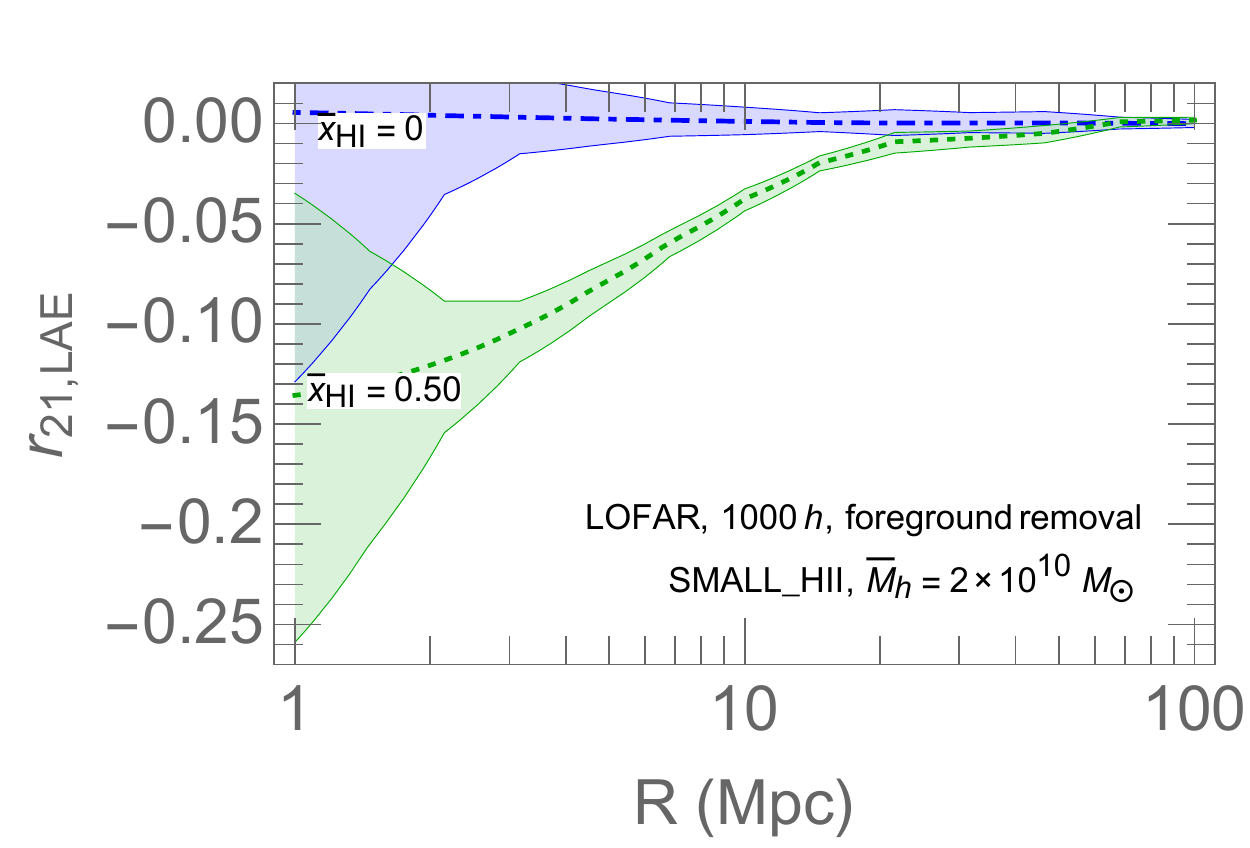}\qquad
\includegraphics[width=0.45\textwidth]{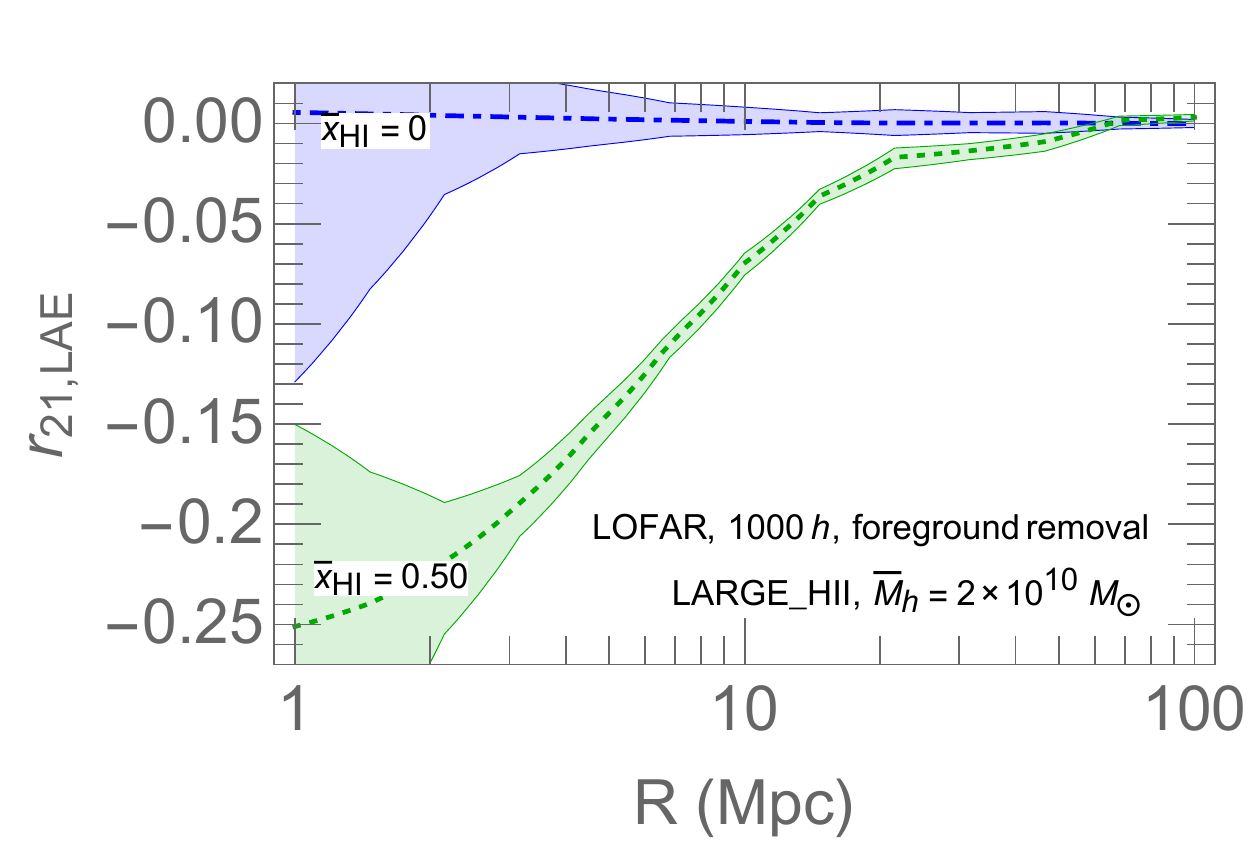}
\includegraphics[width=0.45\textwidth]{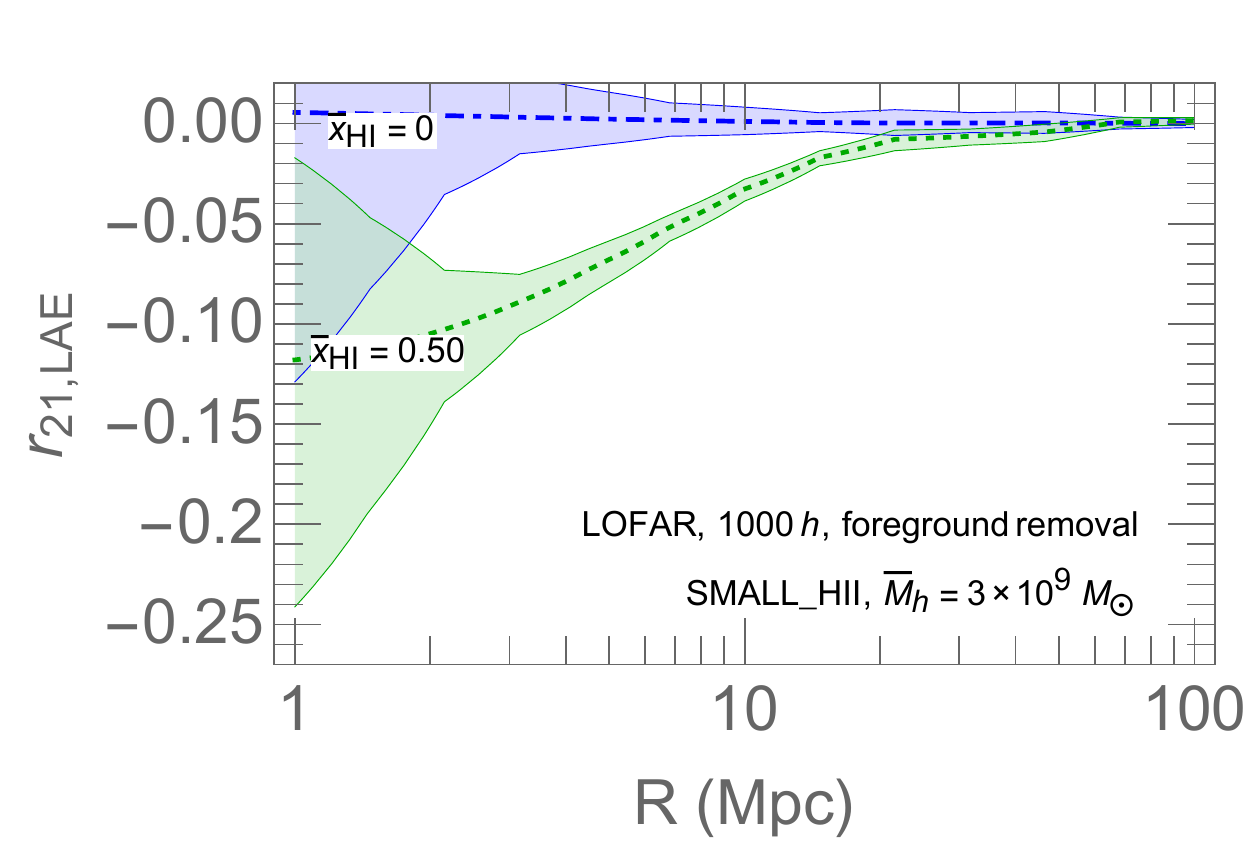}\qquad
\includegraphics[width=0.45\textwidth]{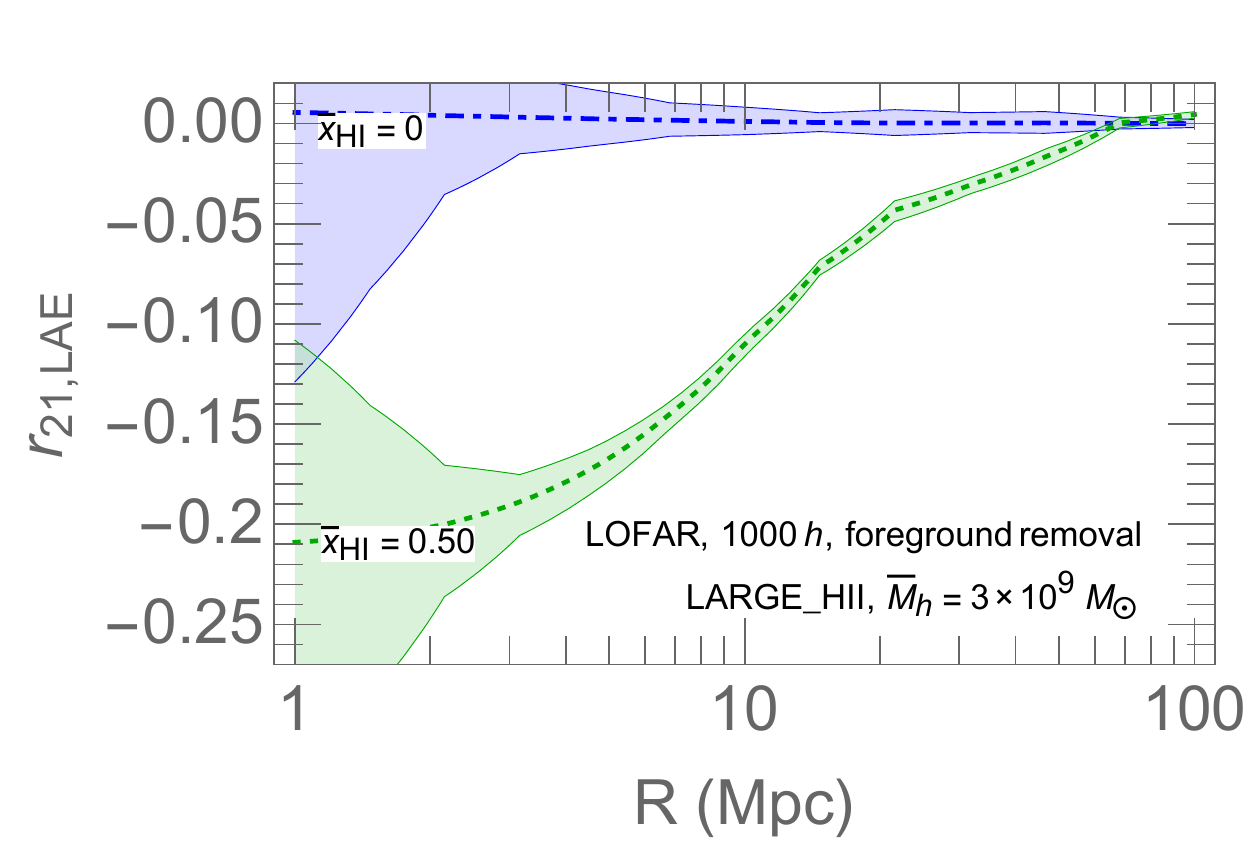}
}
\caption{Same as Fig. \ref{fig:r_lofar_avo}, but assuming optimistic foreground subtraction.
\label{fig:r_lofar_rem}
}
\vspace{-1\baselineskip}
\end{figure*}

\begin{figure*}
\vspace{+0\baselineskip}
{
\includegraphics[width=0.45\textwidth]{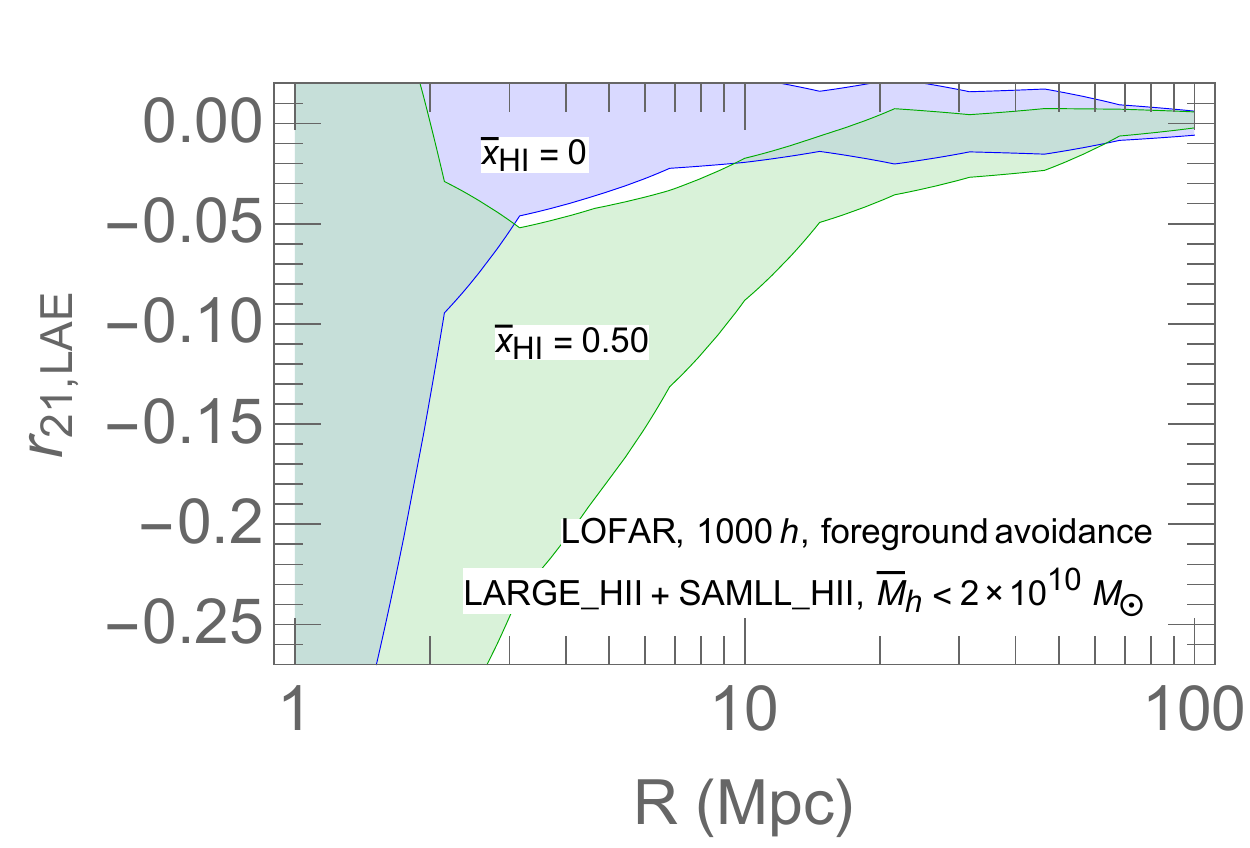}\qquad
\includegraphics[width=0.45\textwidth]{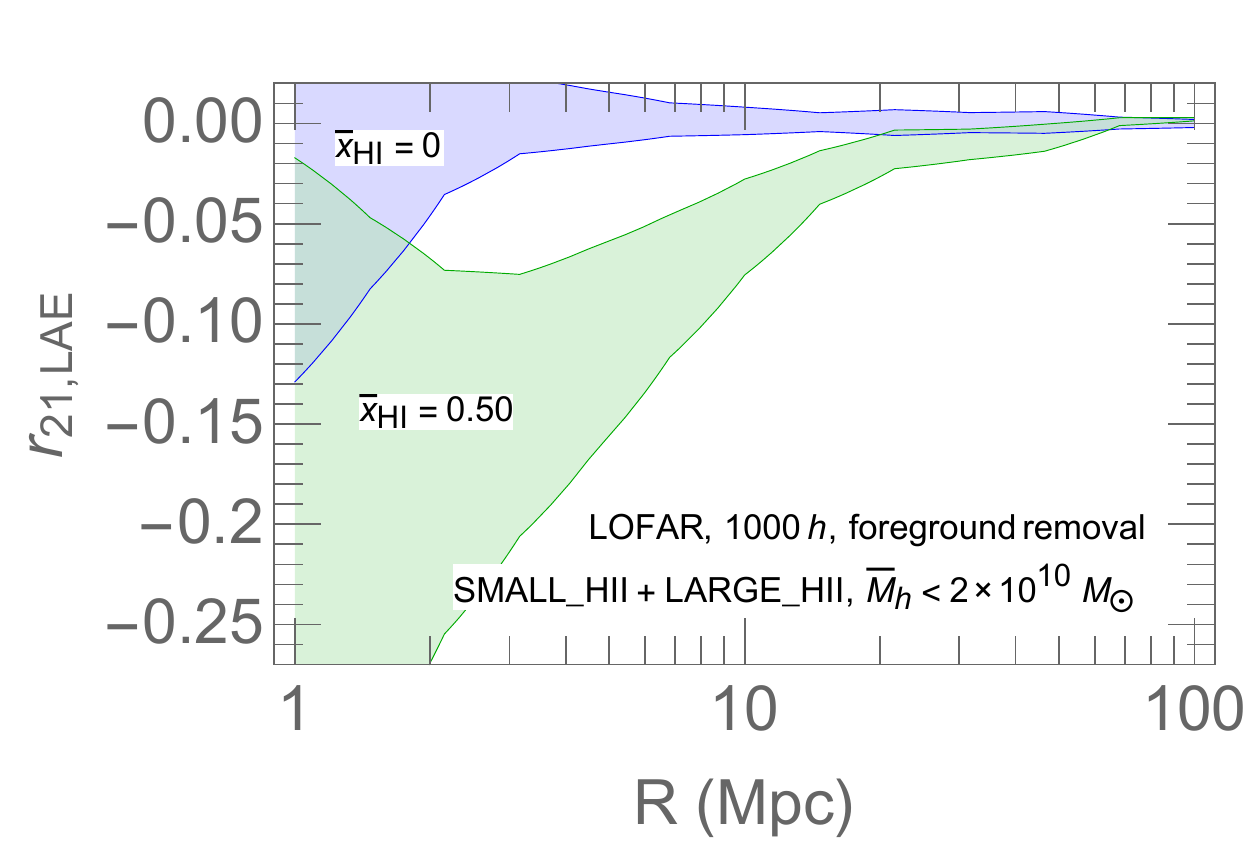}
}
\caption{The LAE-$21\text{ cm}$ cross-correlation function at $\bar{x}_{\rm HI}=0$ (blue) and $\bar{x}_{\rm HI}=0.50$ (green), combining the uncertainties from the previous figures. The shaded regions correspond to total uncertainty due both to the theory (i.e. reionization morphology and average mass $\bar{M}_{\rm h}$ of the halos hosting the LAEs) and the interferometer plus sample noise. The left/right panels correspond to a LOFAR, $1000\text{ h}$ observation, with our conservative/optimistic model for the foregrounds.
\label{fig:LOFAR_main}
}
\vspace{-1\baselineskip}
\end{figure*}

In Figures \ref{fig:r_lofar_avo} and \ref{fig:r_lofar_rem} we show the LAE--21cm brightness temperature cross-correlation function at $\bar{x}_{\rm HI}=0$ (dot-dashed)\footnote{Note that the RSCF at $\bar{x}_{\rm HI}=0$ is weakly positive, since now the dominant 21-cm signal comes from residual HI inside self-shielded systems, which are preferentially near galaxies.  This is in contrast to the 21cm signal during the EoR, in which the HI emitting in 21-cm mostly comes from the large-scale patches of the IGM which have not yet been ionized as they are distant from galaxies.}
  and $\bar{x}_{\rm HI}=0.50$ (dotted). Shaded regions correspond to 1$\sigma$ LOFAR noise for a $1000\text{ h}$ observation, assuming our conservative (``foreground avoidance''; Fig.  \ref{fig:r_lofar_avo}) and  and optimistic (``foreground removal''; Fig. \ref{fig:r_lofar_rem}) model for the foregrounds (see the discussion in \S \ref{sec:obs_21cm}). We show different models for reionization (\textbf{SMALL\pmb{\_}HII}/\textbf{LARGE\pmb{\_}HII} in the left/right panels) and different LAE host halo masses ($\bar{M}_{\rm h}=2\times 10^{10}$/$3\times 10^9M_\odot$ in the upper/lower panels).

Comparing these figures to the left panel of  Fig. \ref{fig:2D},
%LAE--21cm cross-correlations in these figures to the analogous halo--21cm cross-correlations in Fig. \ref{fig:2D}
we see that the LAE--21cm anti-correlation is considerably stronger than the halo--21cm anti-correlation.
This is due to the attenuation of the \lya\ line by the neutral IGM: observed LAEs are more likely to lie deep inside HII regions where the smoothed 21cm signal is the weakest, than a randomly-chosen dark matter halo of the same mass.\footnote{We have tested explicitly that evaluating the \lya\ attenuation at a systemic line offset of $\Delta v_{\rm sys}=-400$ km s$^{-1}$, instead of the fiducial -200 km s$^{-1}$, only impacts the cross-correlation shown in Fig. \ref{fig:r_lofar_avo}-\ref{fig:r_lofar_rem} by $\lesssim 10$\% over the range $R\gsim 1\text{ Mpc}$.
This is consistent with the results of \citet{SM15} (see their Fig. A1 and associated discussion), which showed that the angular LAE correlation functions during the EoR are very insensitive to the intrinsic emission line, when normalized to the same number density and typical halo mass.}  However, the same trends from Fig. \ref{fig:2D} are evident.  Namely, the turnover (defined as the scale on which the RSCF is half of its smallest shown value) occurs on the same scale, which is a function of the characteristic HII region scale.

Moreover, while the EoR morphology has a strong impact on the cross-correlation (comparing left and right panels), the LAE model does not (comparing top and bottom panels). This is in contrast to the observed LAE angular correlation function, which does depend on the intrinsic host halos \citep{McQuinn07LAE, SM15}.  Thus the LAE--21cm cross correlation is a robust probe of the EoR, {\it insensitive to uncertainties in LAE modeling}.

In Figure \ref{fig:LOFAR_main} we combine the theoretical uncertainties from the EoR morphology and the LAE models spanned by our models.
Thus, the shaded regions correspond to cumulative uncertainty due both to the theory (i.e. reionization morphology and average mass $\bar{M}_{\rm h}$ of the halos hosting the LAEs) and the interferometer noise plus sample variance (the left/right panels correspond to a LOFAR, $1000\text{ h}$ observation, with our conservative/optimistic model for the foregrounds).  We see that the purple and green shaded regions do not overlap at scales of $R\sim$ 3--10 Mpc.  This implies that a fully ionized and half ionized Universe can be distinguished using a 1000h observation with LOFAR correlated with the HSC UDF, even assuming maximally pessimistic models for the EoR morphology with small HII regions which minimize the cross-correlation signal.  This detection can be made with a S/N of 1--2, depending on the foreground model.

\subsubsection{SKA1-Low}

\begin{figure*}
\vspace{+0\baselineskip}
{
\includegraphics[width=0.45\textwidth]{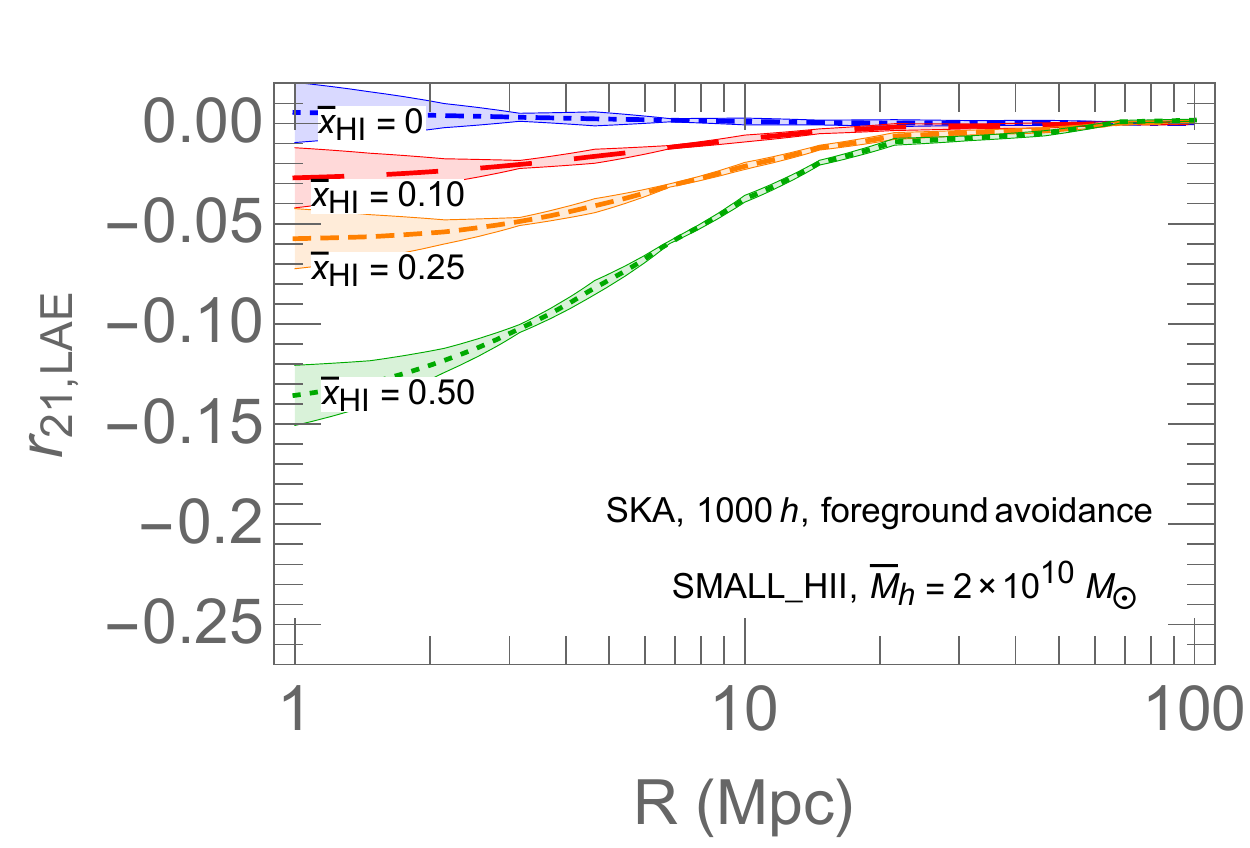}\qquad
\includegraphics[width=0.45\textwidth]{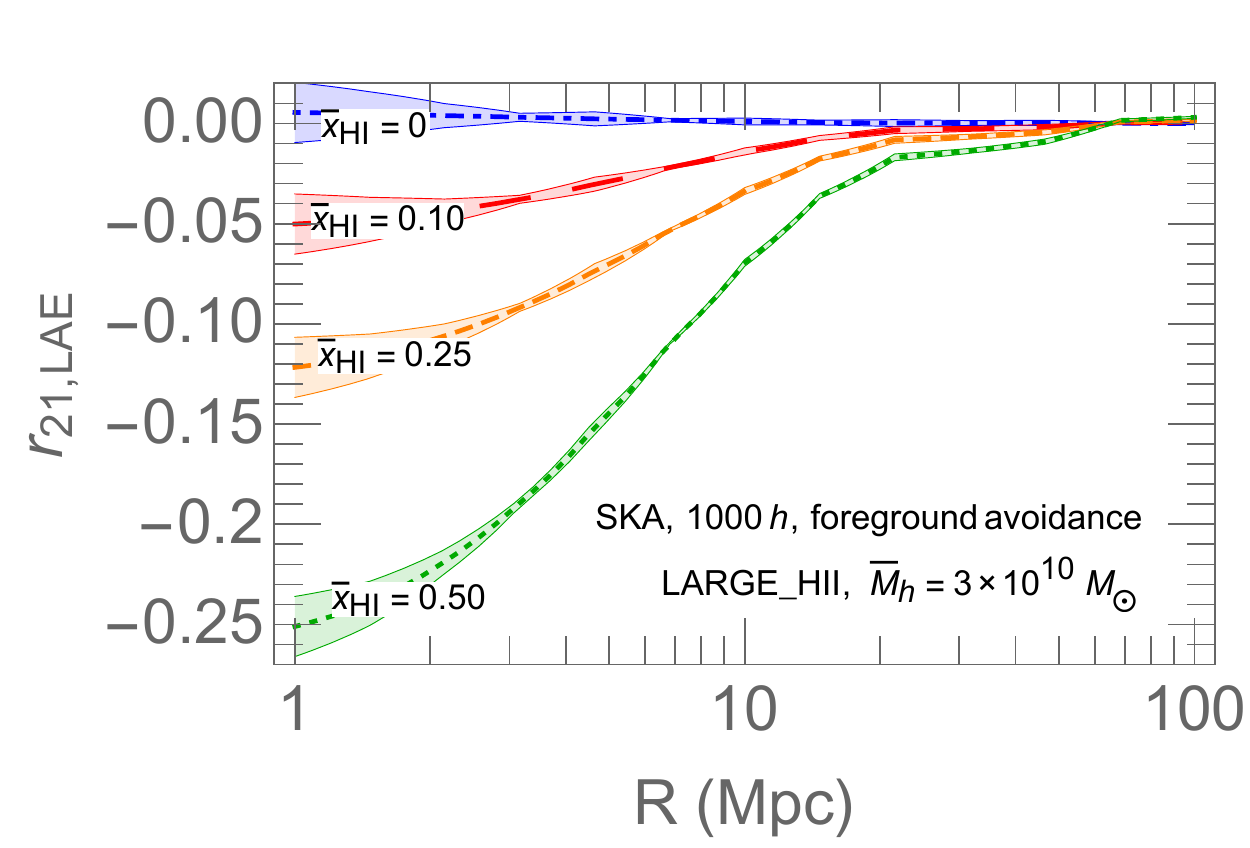}
\includegraphics[width=0.45\textwidth]{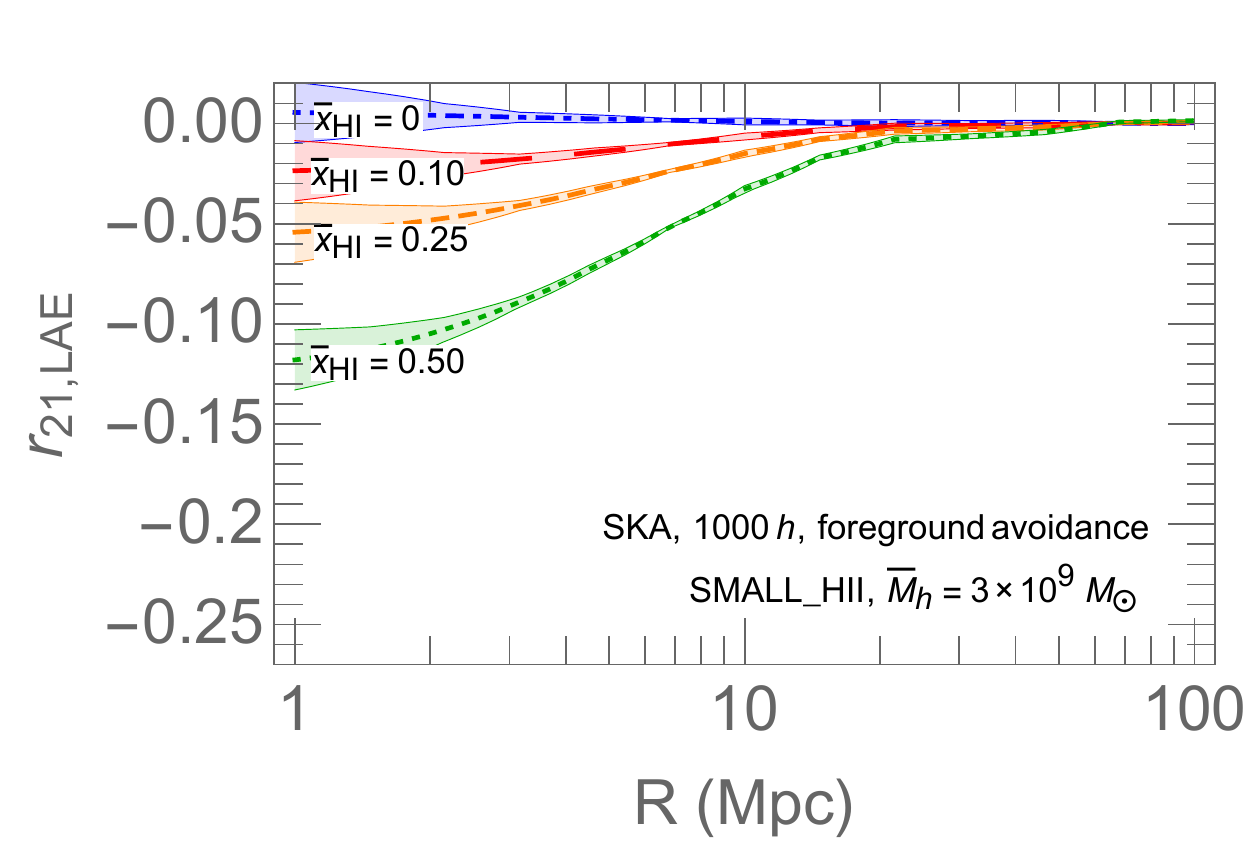}\qquad
\includegraphics[width=0.45\textwidth]{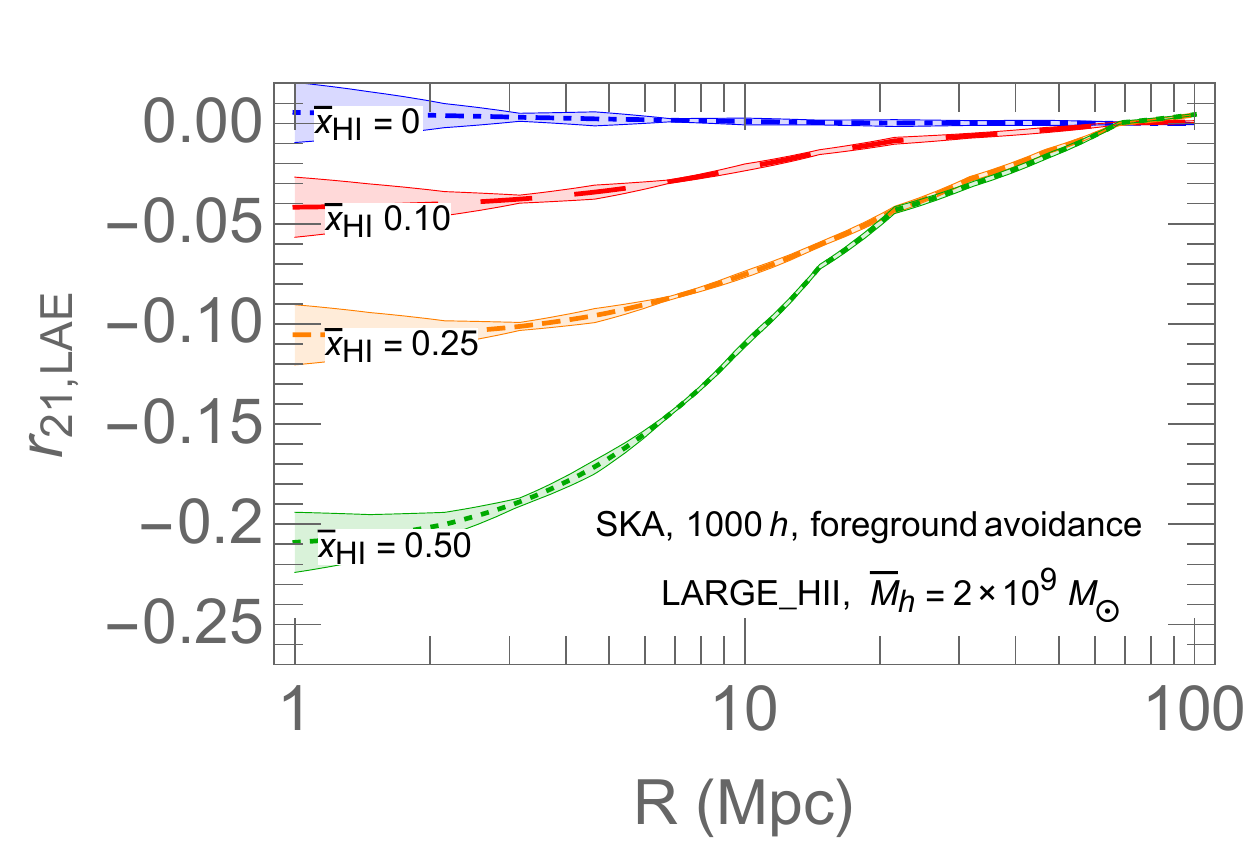}
}
\caption{Same as Fig.  \ref{fig:r_lofar_avo}, but computed for a 1000 h observation with SKA1-Low.
\label{fig:r_ska}
}
\vspace{-1\baselineskip}
\end{figure*}

\begin{figure*}
\vspace{+0\baselineskip}
{
\includegraphics[width=0.45\textwidth]{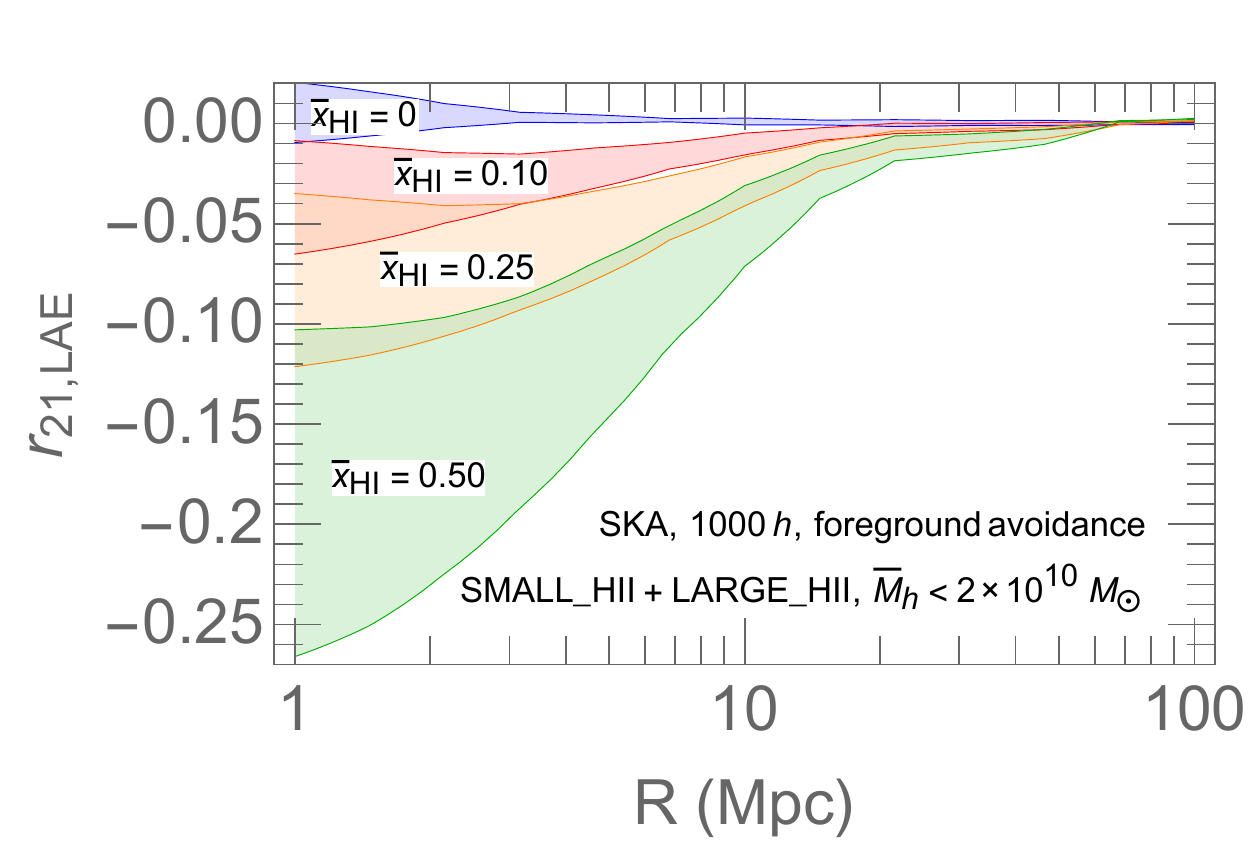}\qquad
\includegraphics[width=0.45\textwidth]{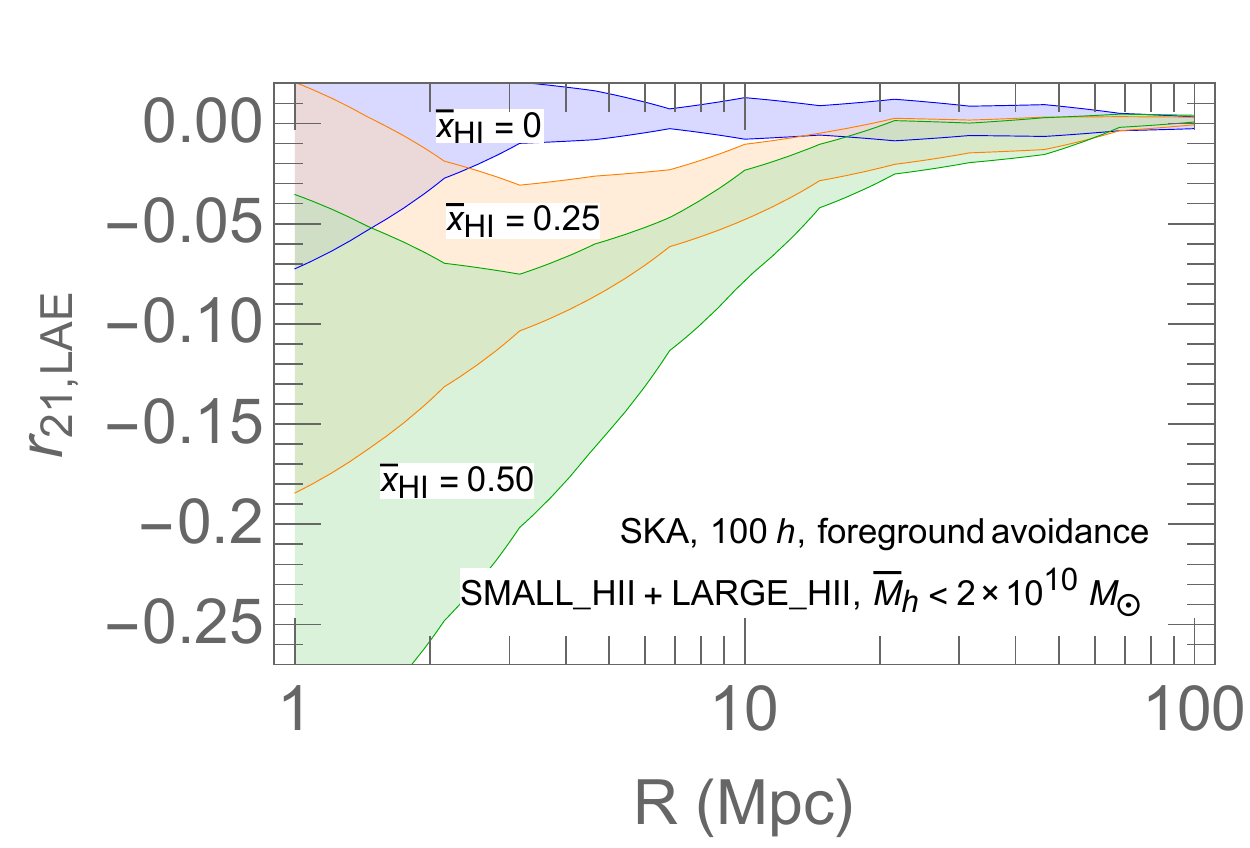}
}
\caption{The LAE-$21\text{ cm}$ brightness temperature cross-correlation function at $\bar{x}_{\rm HI}=0$ (blue), $\bar{x}_{\rm HI}=0.10$ (red), $\bar{x}_{\rm HI}=0.25$ (orange) and $\bar{x}_{\rm HI}=0.50$ (green). The shaded regions correspond to total uncertainty due both to the theory (i.e. reionization morphology and average mass $\bar{M}_{\rm h}$ of the halos hosting the LAEs) and the interferometer noise plus sample variance. The left/right panels correspond to an SKA1-Low, $1000\text{ h}$/$100\text{ h}$ observation, assuming foreground avoidance.
\label{fig:SKA_main}
}
\vspace{-1\baselineskip}
\end{figure*}

In Figure \ref{fig:r_ska} we show the LAE-$21\text{ cm}$ cross-correlation function at $\bar{x}_{\rm HI}=0$, 0.1, 0.25, 0.5.
Shaded regions correspond to SKA1-Low noise for a $1000\text{ h}$ observation.  Since the SKA noise is much smaller than the uncertainty from the EoR morphology, we only adopt the conservative model for the foregrounds.
As for LOFAR, we show different models for reionization (\textbf{SMALL\pmb{\_}HII}/\textbf{LARGE\pmb{\_}HII} in the left/right panels) and for the host halo masses ($\bar{M}_{\rm h}=2\times 10^{10}$/$3\times 10^9M_\odot$ in the upper/lower panels).
The noise is reduced by factors of $> 10$ with respect to LOFAR, allowing the cross-correlation to discriminate easily between neutral fractions which are different by $\sim$ per cent.

Mimicking the previous LOFAR analysis, in Figure \ref{fig:SKA_main} we combine the theoretical uncertainties in the EoR modeling, which now dominate over the interferometer noise.  From the left panel, we see that a $1000\text{ h}$ SKA observation can discriminate between a completely ionized Universe from one with $\bar{x}_{\rm HI}=0.10$.  Or alternately, it can discriminate $\bar{x}_{\rm HI}=0.10$ from $\bar{x}_{\rm HI}=0.50$.

However, it is unrealistic to expect SKA to devote a deep, 1000h observation to the same field as Subaru, given that the overlap of the two FoV is not ideal.
A more realistic scenario would take advantage of the planned 100h SKA survey \citep{Koopmans15}, whose wider area could more easily accommodate an overlap with a Subaru field.  The LAE--21cm cross correlation within the area of overlap can then be used as a test of SKA's data analysis pipelines.

In the right panel of Figure \ref{fig:SKA_main}, we show the analogous signal as in the left panel, but computed assuming only 100h of integration.  We see that even with the corresponding increase in noise, the $100\text{ h}$ observation can discriminate between a fully ionized Universe and one with $\bar{x}_{\rm HI}=0.25$.  Note also that the uncertainty for these forecasts is dominated by our lack of knowledge about the EoR morphology;
thus these results can be significantly improved with a model prior on the allowed EoR morphology from theory or complimentary observations.

%%%%%%%%%%%%%%%%%%%%%%%%%%%%%%%%%%%%%%%%%%%%%%%%%%%%%%%%%%%%%%%%%%%%%%%%%%%%%%%%%%%%%%%%%%%%%%%%%%%%%%%%%%%%%%%%%%%%%%%%%%%%%%%%%%%%%%%%
%%%%%%%%%%%%%%%%%%%%%%%%%%%%%%%%%%%%%%%%%%%%%%%%%%%%%%%%%%%%%%%%%%%%%%%%%%%%%%%%%%%%%%%%%%%%%%%%%%%%%%%%%%%%%%%%%%%%%%%%%%%%%%%%%%%%%%%%
\section{Conclusions}
\label{sec:concl}

Ongoing and upcoming efforts at 21-cm cosmology face significant challenges in dealing with systematics.  Since systematics should not correlate with genuine cosmological signals, observing such a correlation would lend credibility to any putative claims of an EoR detection.  Here we present forecasts for the cross-correlation of the 21-cm signal with upcoming wide-field LAE surveys with the Subaru HSC.

We study the dependence of the LAE-21cm  correlation function on the average halo mass hosting LAEs and on the EoR morphology. The RSCF is very insensitive to the intrinsic clustering of LAEs, making it a robust probe of the EoR.  Different EoR morphologies change the value of the RSCF by up to a factor of $\sim 2$ at a given scale.

We present forecasts for the LAE-21cm cross-correlation for LOFAR and SKA1-Low.  A $1000\text{ h}$ LOFAR observation can discriminate a fully ionized Universe from $\bar{x}_{\rm HI}=0.50$ looking at the RSCF at scales of $3-10\text{ Mpc}$. The significance of this detection is limited by LOFAR noise.

On the other hand for SKA1-Low, the main limitation is our ignorance of the EoR morphology.  However, even with maximally pessimistic assumptions, a $1000\text{ h}$ observation with SKA1-Low can discriminate $\bar{x}_{\rm HI}=0$ from $\bar{x}_{\rm HI}=0.10$ and $\bar{x}_{\rm HI}=0.10$ from $\bar{x}_{\rm HI}=0.50$.

More practical however would be to cross-correlate the LAE maps with a shallower, wider SKA1-Low survey, using the resulting detection as a sanity check on data analysis efforts.  Indeed, we find that the LAE-21cm cross-correlation from the planned 100 h SKA1 survey is sufficient to discriminate a fully ionized Universe from $\bar{x}_{\rm HI}=0.25$.  Priors on the EoR morphology (from either theory or complementary observations) can substantially improve the significance of this detection.

\section*{Acknowledgements}

This project has received funding from the European Research Council (ERC) under the European Union's Horizon 2020 research and innovation programme (grant agreement No 638809 -- AIDA).

\bibliographystyle{mn2e}
\bibliography{ms}

\end{document}